\documentclass{aa}  
\usepackage{graphicx}
\usepackage{txfonts}
\usepackage{ulem}
\usepackage{sidecap}
\begin{document} 
   \title{Solar Atmosphere Radiative Transfer Model Comparison based on 3D MHD simulations}

   \author{M. Haberreiter
	\inst{1}
	\and
	S. Criscuoli
	\inst{2}
	\and
	M. Rempel
	\inst{3}           
	\and
	T.M.D. Pereira
	\inst{4, 5} 
}
\institute{Physikalisch-Meteorologisches Observatorium Davos and World Radiation Center, Dorfstrasse 33, 7260 Davos Dorf, Switzerland\\
	\email{margit.haberreiter@pmodwrc.ch}
	\and
	National Solar Observatory, 3665 Discovery Dr., Boulder, CO 80303, USA
	\and
	High Altitude Observatory, NCAR, P.O. Box 3000, Boulder, CO 80307, USA
	\and	
	Institute of Theoretical Astrophysics, University of Oslo, P.O. Box 1029 Blindern, 0315 Oslo, Norway
	\and
	Rosseland Centre for Solar Physics, University of Oslo, P.O. Box 1029 Blindern, 0315 Oslo, Norway
}
   \date{Accepted: 26 May 2021}

 
  \abstract
   {The reconstruction of the solar spectral irradiance (SSI) on various time scales is essential for the understanding of the Earth's climate response to the SSI variability. }
   {The driver of the SSI variability is understood to be the intensity contrast of magnetic features present on the Sun with respect to the largely non-magnetic quiet Sun. However, different spectral synthesis codes lead to diverging projections of SSI variability. In this study we compare three different radiative transfer codes and carry out a detailed analysis of their performance.}
   {We perform the spectral synthesis at the continuum wavelength of 665 nm with the { Code for Solar Irradiance (COSI), and the Rybicki-Hummer (RH), and Max Planck University of Chicago Radiative MHD (MURaM)} codes for three 3D MHD simulations snapshots, a non-magnetic case, and MHD simulations with 100 G, and 200 G magnetic field strength.}
   {We determine the intensity distributions, the intensity differences and ratios for the spectral synthesis codes. We identify that the largest discrepancies originate in the intergranular lanes where the most field concentration occurs. }
   {Overall, the applied radiative transfer codes give consistent intensity distributions. Also, the intensity variation as a function of magnetic field strength for the particular 100 G and 200 G snapshots agree within the 2-3\% range.}

   \maketitle
%
\section{Introduction}
Solar irradiance varies on short timescales from minutes to hours as well as long time scales of days, years, decades and beyond. The variations at the different time scales originate from different processes in the solar atmosphere. The short time scales - other than reconnection processes such as flares - are mostly determined by convection and the longer ones are largely driven by the changing solar surface magnetic field \citep[see e.g., ][]{Domingo2009,Solanki2013ARAA,yeo2017b}. 

The results over the past decades clearly indicate that solar variability has an influence on the Earth's climate; for an overview see e.g. \cite{Matthes2017} and \cite{Shindell2020}. However, the exact quantification of the solar influence on climate - besides other natural forcings and the anthropogenic contribution - is still debated. \cite{Egorova2018FrEaS} conclude that the Sun must have varied substantially, in order to attribute the temperature increase in the early 19th century. 

To quantify the solar contribution to climate change more precisely, robust solar irradiance datasets are needed as input to the climate models. For those times when space observations provide a decent temporal and spectral coverage, a number of observational composite datasets are available \citep[see e.g., ][]{Haberreiter2017,Coddington2019,Marchenko2019}. However, when no satellite observations are available irradiance reconstruction models using proxy datasets to describe the state of solar activity back in time are key for our understanding of its impact on climate.

The extent to which solar variability has changed over long time scales is still an open question. \cite{Shapiro2011} proposed a reconstruction of the {Total Solar Irradiance (TSI) and Spectral Solar Irradiance (SSI)}, which accounts for a variable quiet Sun intensity. This approach leads to a significant change of the radiative forcing between the Maunder Minimum and the space era of about 6$\pm$3 W m$^{-2}$. A later update of this approach suggests a smaller change in long-term irradiance variability \citep{Egorova2018AA}. However, other reconstructions methods give a rather flat long-term trend; see e.g. \cite{Solanki2013ARAA}. Recently, {\cite{Rempel2020}} determined a linear dependence between the outgoing radiative energy flux and the mean magnetic field strength for the quiet Sun. {While the strong sensitivity of TSI to the quiet Sun field strength implies that potential variations of the magnetic field over longer timescales could make a significant contribution to solar irradiance variations, such magnetic variations are not expected if the quiet Sun magnetic field originates primarily from a small-scale dynamo}. 
\begin{figure*}[th!]
\begin{center}
	\includegraphics[width=0.9\linewidth, angle=0]{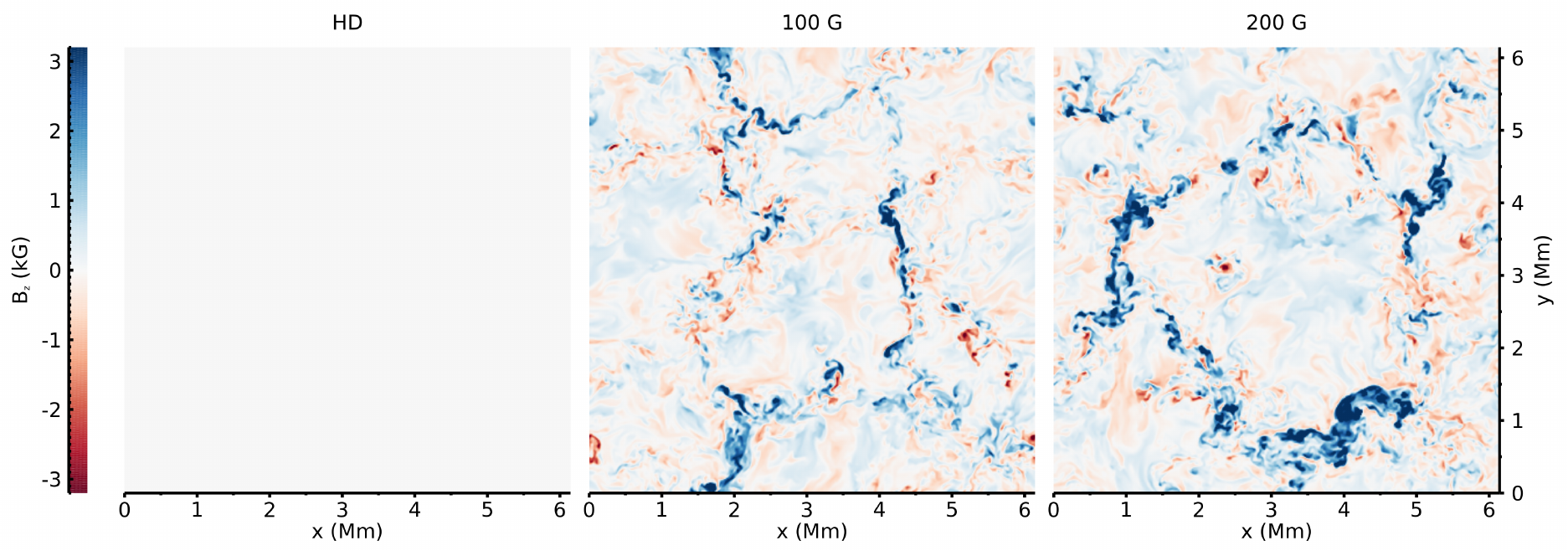}\\
	\hspace*{0.cm}\includegraphics[width=0.9\linewidth]{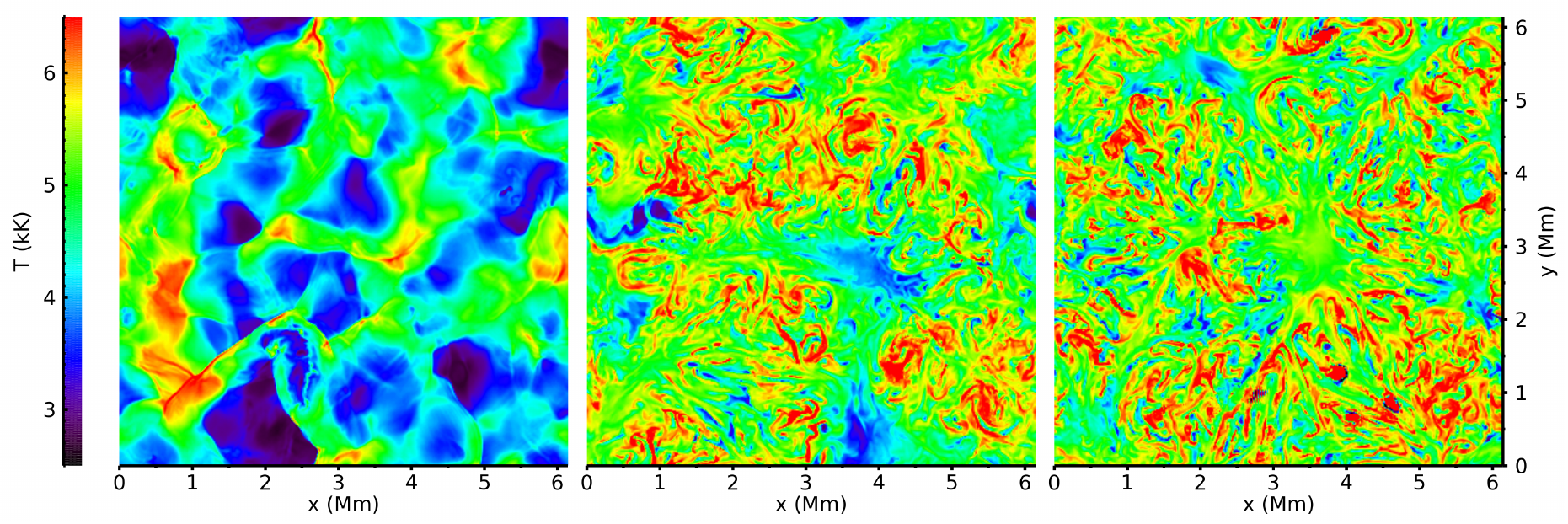} 
\caption{\label{fig:magz} Top panels: {Photospheric} magnetic field strength $B_{z,\tau=1}$ for the snapshots of the pure hydrodynamic case (left panel), and the cases with a magnetic field strength of 100 G (middle panel), and 200 G (right panel); bottom panels: Horizontal temperature variation of the snapshots 
{at the top of the simulation box} for the pure hydrodynamical case (left panel), and the cases with a magnetic field strength of 100 G (middle panel), and 200 G (right panel).}
\end{center}
\end{figure*}
Nevertheless, to determine the irradiance variations correctly, it is key to quantify the emergent spectrum for various solar surface magnetic elements with a radiative transfer code. State-of-the-art radiative transfer codes are however inherently different with respect to the atomic input data and {different numerical schemes to solve the radiative transfer equation}. This introduces some uncertainty for irradiance reconstructions. \cite{Criscuoli2020} recently validated the performance of a set of commonly used codes, i.e. the radiative transfer codes COSI \citep{Haberreiter2008b,shapiro2010,Criscuoli2020}, the RH code \citep{Uitenbroek2003}, and the radiation scheme of the MURaM code \citep{Rempel2014}. The authors find a good agreement of the spectral synthesis using 1D solar atmosphere structures.
In the present paper we go beyond the study by \cite{Criscuoli2020} and use the vertical temperature and density profiles from three 3D MHD simulations, a hydrodynamic (HD) case with 0-G magnetic field, and a 100-G and 200-G MHD snapshot as input for the spectral synthesis to the above-mentioned radiative transfer codes in order to {assess} their performance. This is deemed particularly necessary, as radiative transfer codes developed to work in 3D geometry may differ not only for the reasons mentioned above, but also for the numerical schemes employed to resolve discontinuities and strong gradients typically present in atmospheres generated by MHD codes \citep[see e.g.][]{Janett2019}.   
Similar work has already been carried out by \cite{afram2011} and \cite{Norris2017} by applying a set of simulations with different magnetic field strengths from the MURaM Code \citep{Voegler2005} as input for the spectral synthesis. To our knowledge, this is the first time that {spectral syntheses from 3D MHD simulations carried out by different radiative transfer codes are compared in a quantitative manner.}

The paper is organized as follows. First, in Sec.\,\ref{sec:mhd} we introduce the 3D MHD simulations used to calculate the synthetic spectra. Second, in Sec.\,\ref{sec:codes} we describe the radiative transfer codes and in Sec.\,\ref{sec:results} we discuss the results. 
Finally, in Sec.\,\ref{sec:concl} we summarize our findings.
\begin{figure*}[th!]
\sidecaptionvpos{}{}
	\begin{center}
    \includegraphics[width=0.33\linewidth]{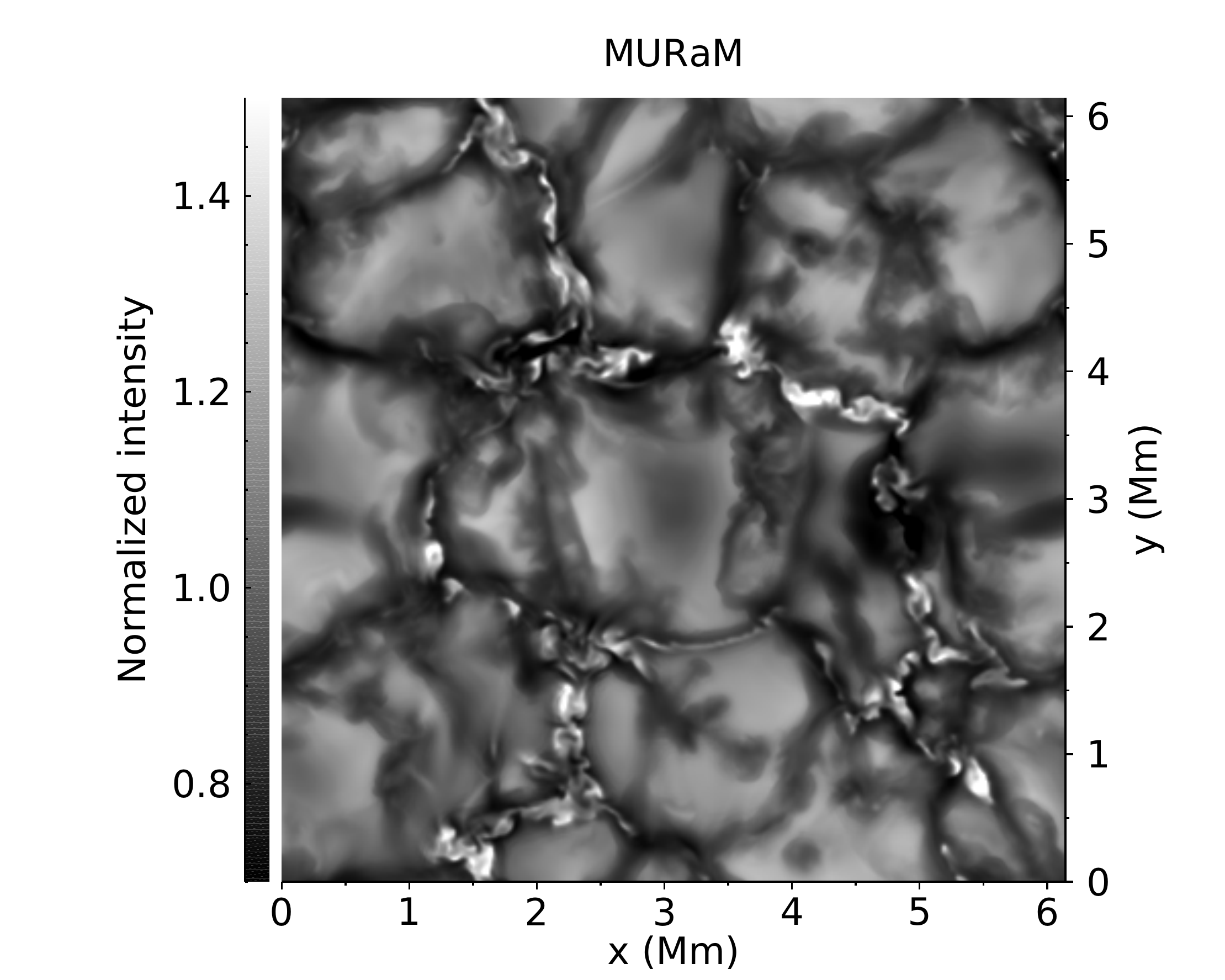} 
    \includegraphics[width=0.33\linewidth]{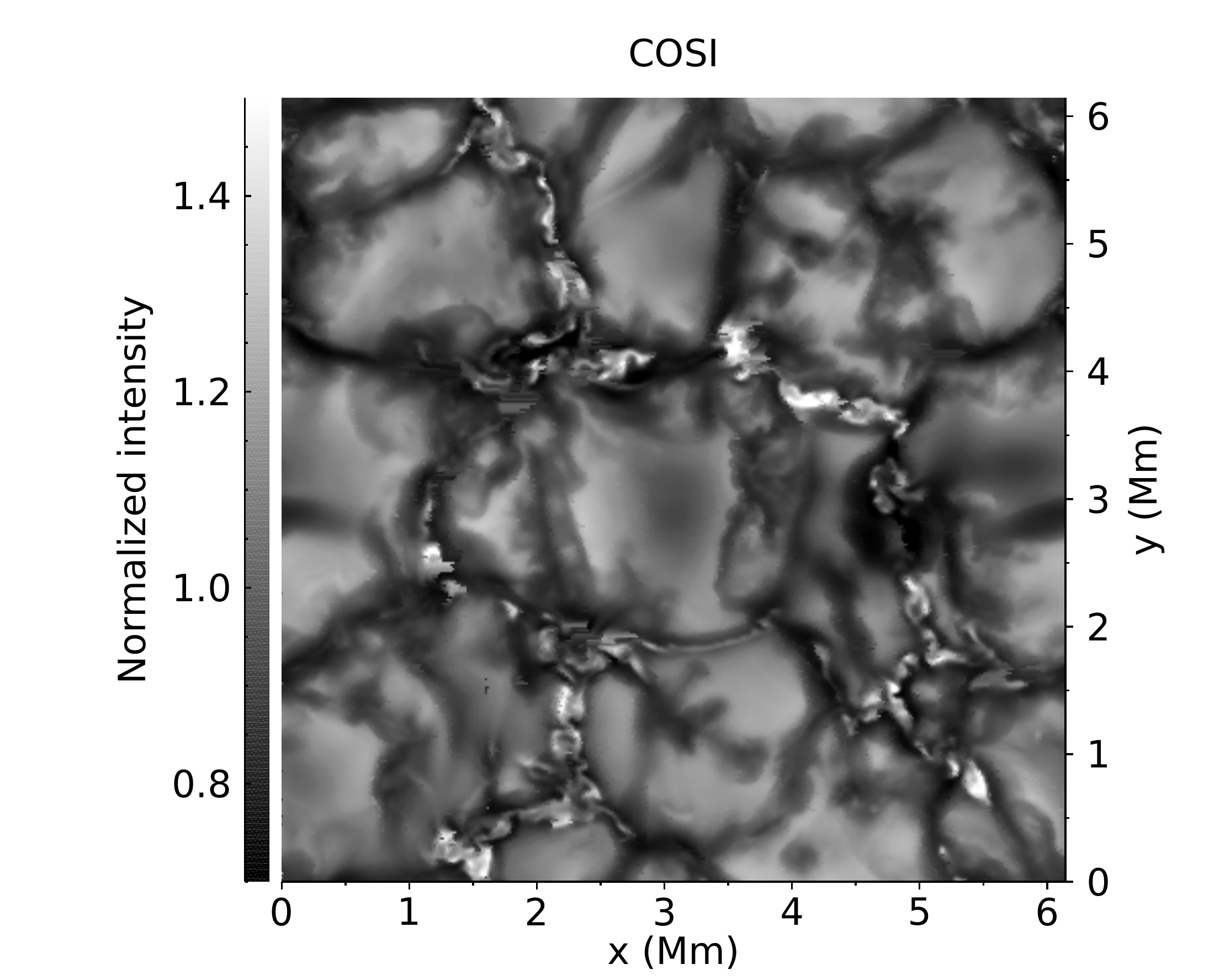}\\
    \includegraphics[width=0.33\linewidth]{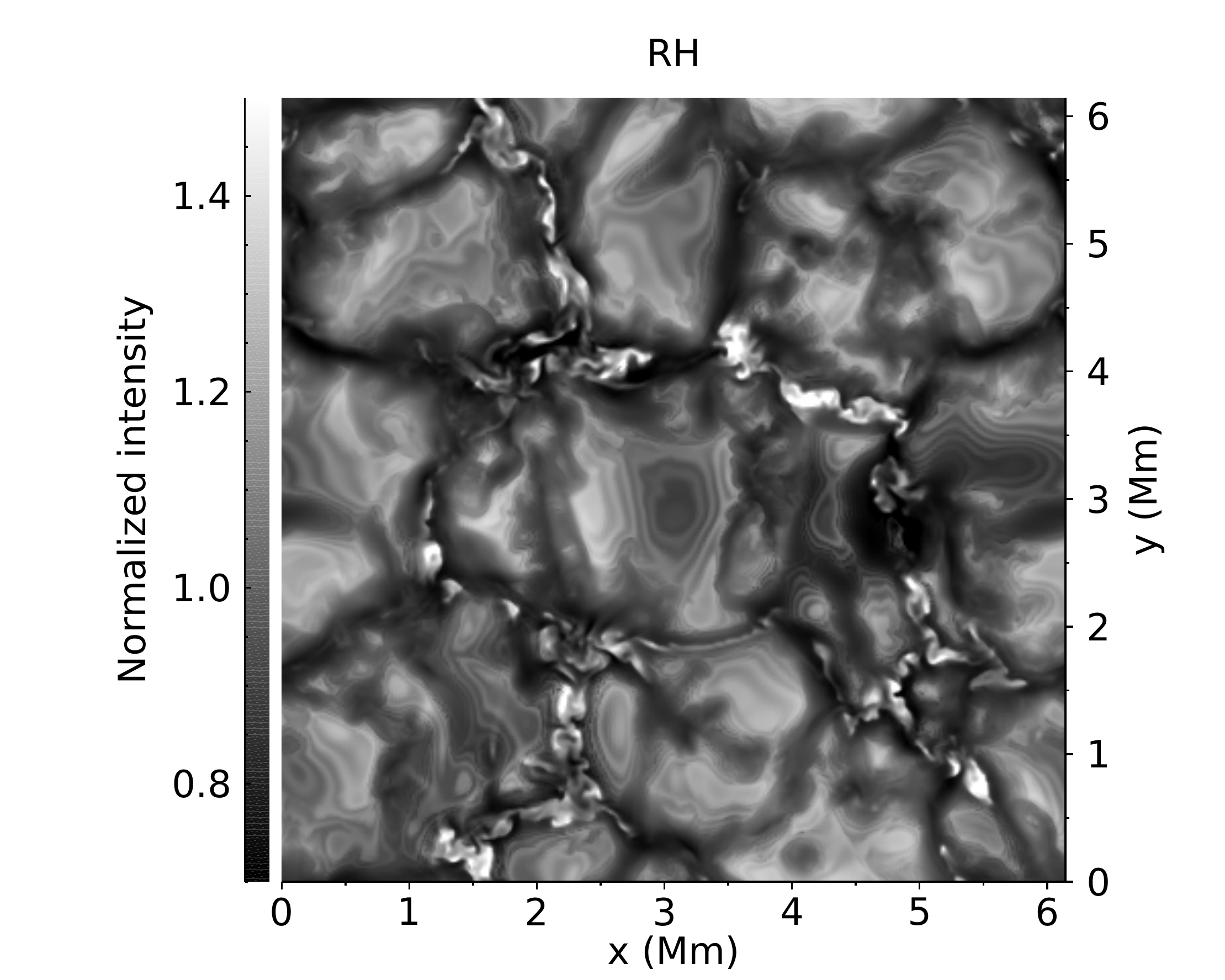}
    \includegraphics[width=0.33\linewidth]{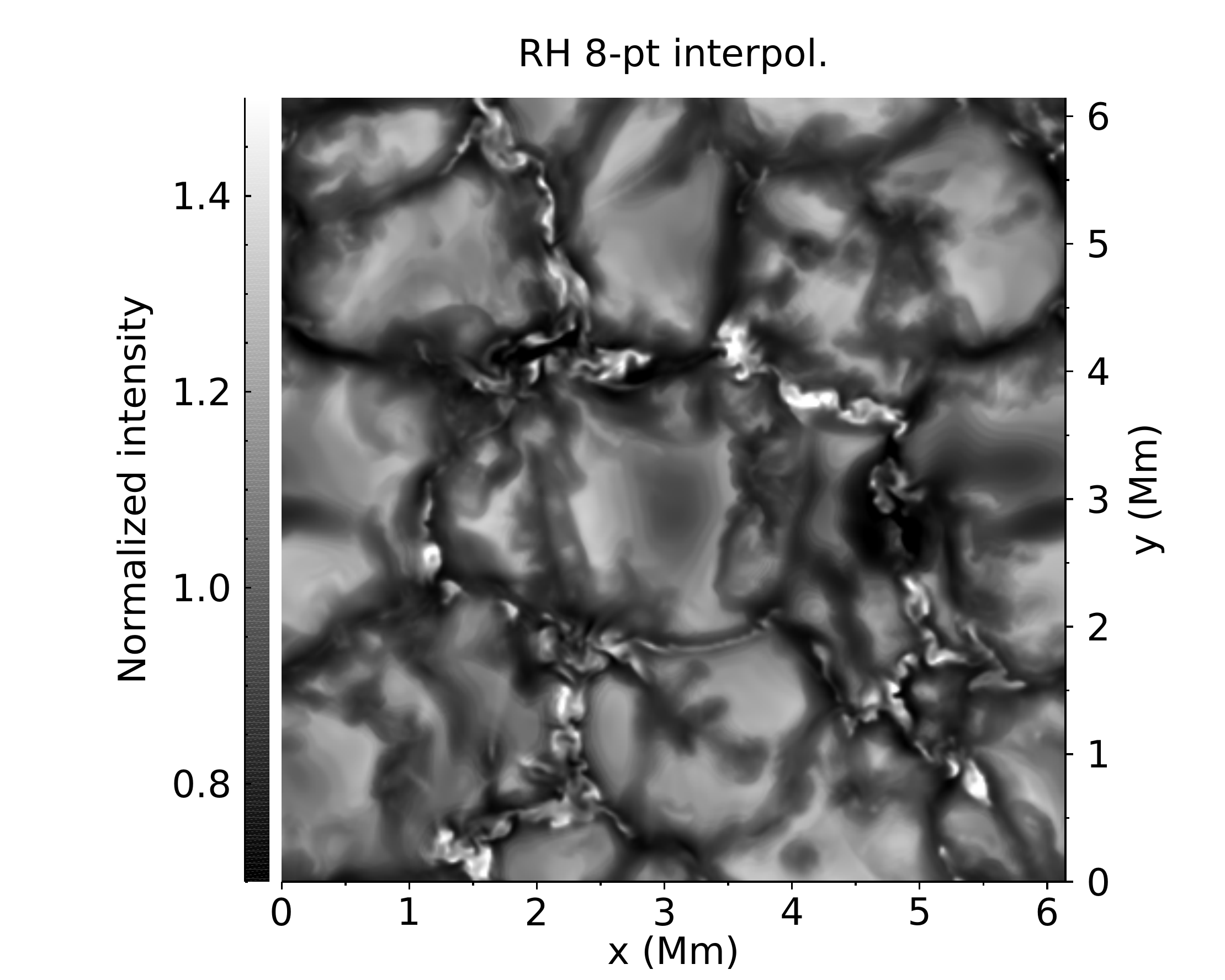}
\caption{\label{fig:fringes} Relative intensity calculated with the MURaM (top left), COSI (top right), RH based on the original MHD grid (bottom left), and RH based on the 8-pt interpolated MHD grid (bottom right). Each snaphot was normalized to its respective mean intensity.}
	\end{center}
\end{figure*}
\section{MHD simulations}\label{sec:mhd}
For this study we calculate the emergent intensities for {different} snapshots from the simulation runs with the MURaM code \citep{Rempel2014} for the non-magnetic, or hydrodynamic (HD) case, and for a magnetic field of $100$~G and $200$~G, respectively. The magnetic cases were branched from the HD setup by adding a uniform vertical field of 100 and 200 G. The simulations ran for about an hour {simulated solar time} until a statistically relaxed state was reached. These simulations consist of cubes of $384\times384\times96$~pixels$^3$, corresponding to an area on the solar surface of $8.8 \times 8.8$ arcsec$^2$. {The vertical domain extent is 1536 km, with the average tau=1 level located at about 684 km beneath the top boundary}.
For further details we refer to  \cite{Criscuoli2020}.
Figure\,\ref{fig:magz}, top panels, show the horizontal distribution of the line-of-sight magnetic field strength at the $\tau$=1 layer, and the temperature variation at the top of the simulation box. The bottom panels give an example of the horizontal variation of the temperature. The {depth stratification of the} temperature and density are used as input to the calculations discussed in Sec.\,\ref{sec:codes}.
\begin{figure*}[th!]
	\begin{center}
		\includegraphics[width=0.33\linewidth]{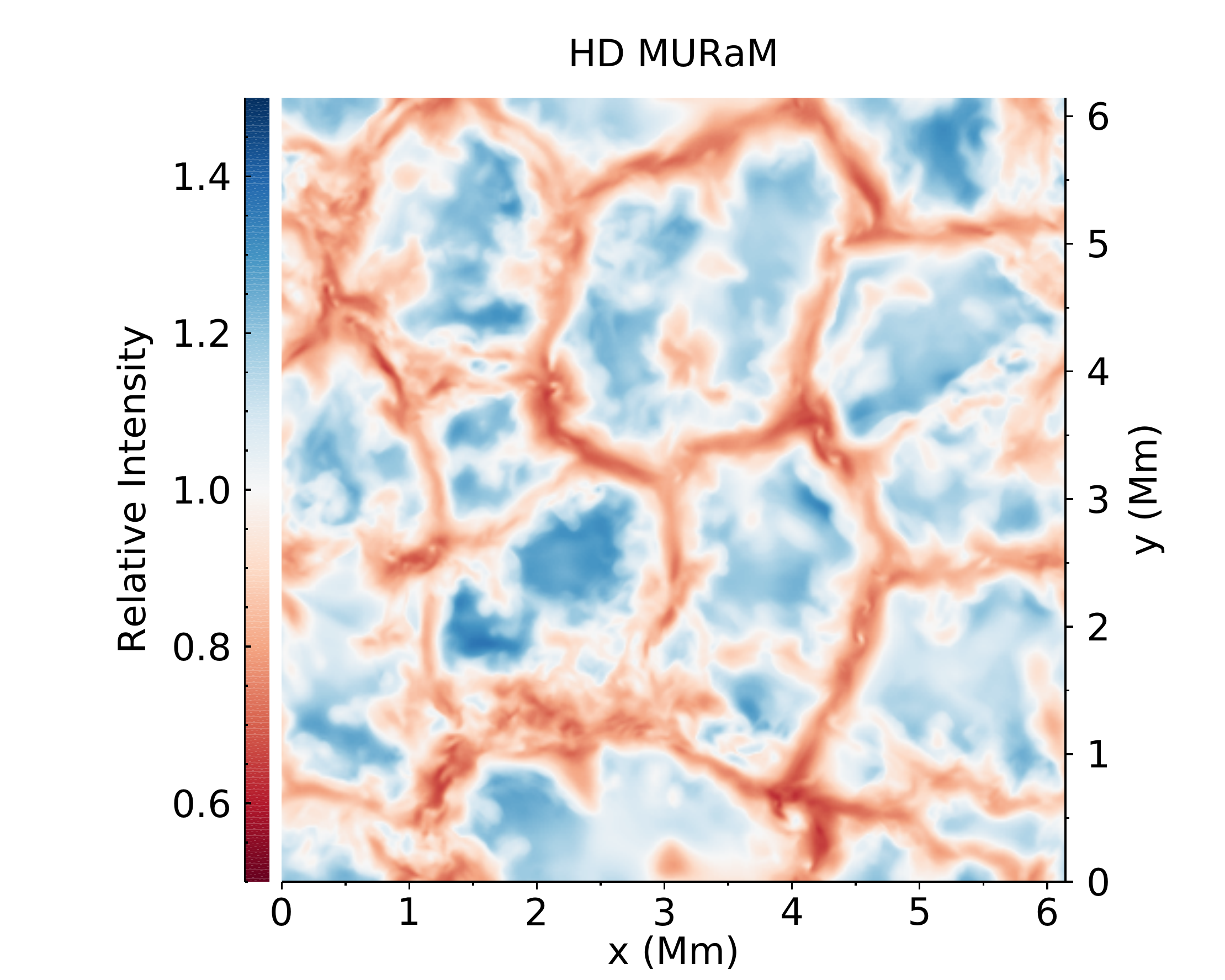} 
		\includegraphics[width=0.33\linewidth]{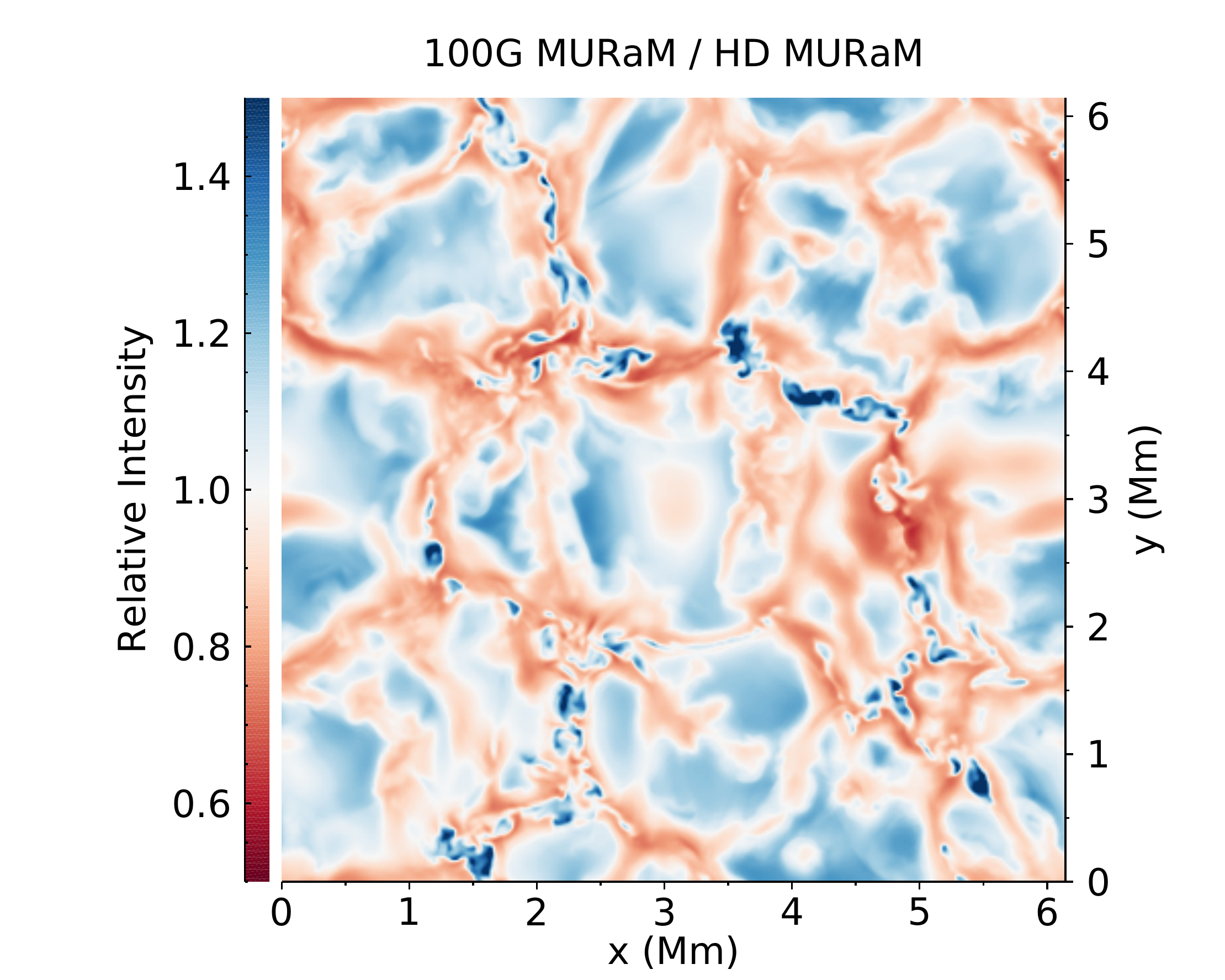}
		\includegraphics[width=0.33\linewidth]{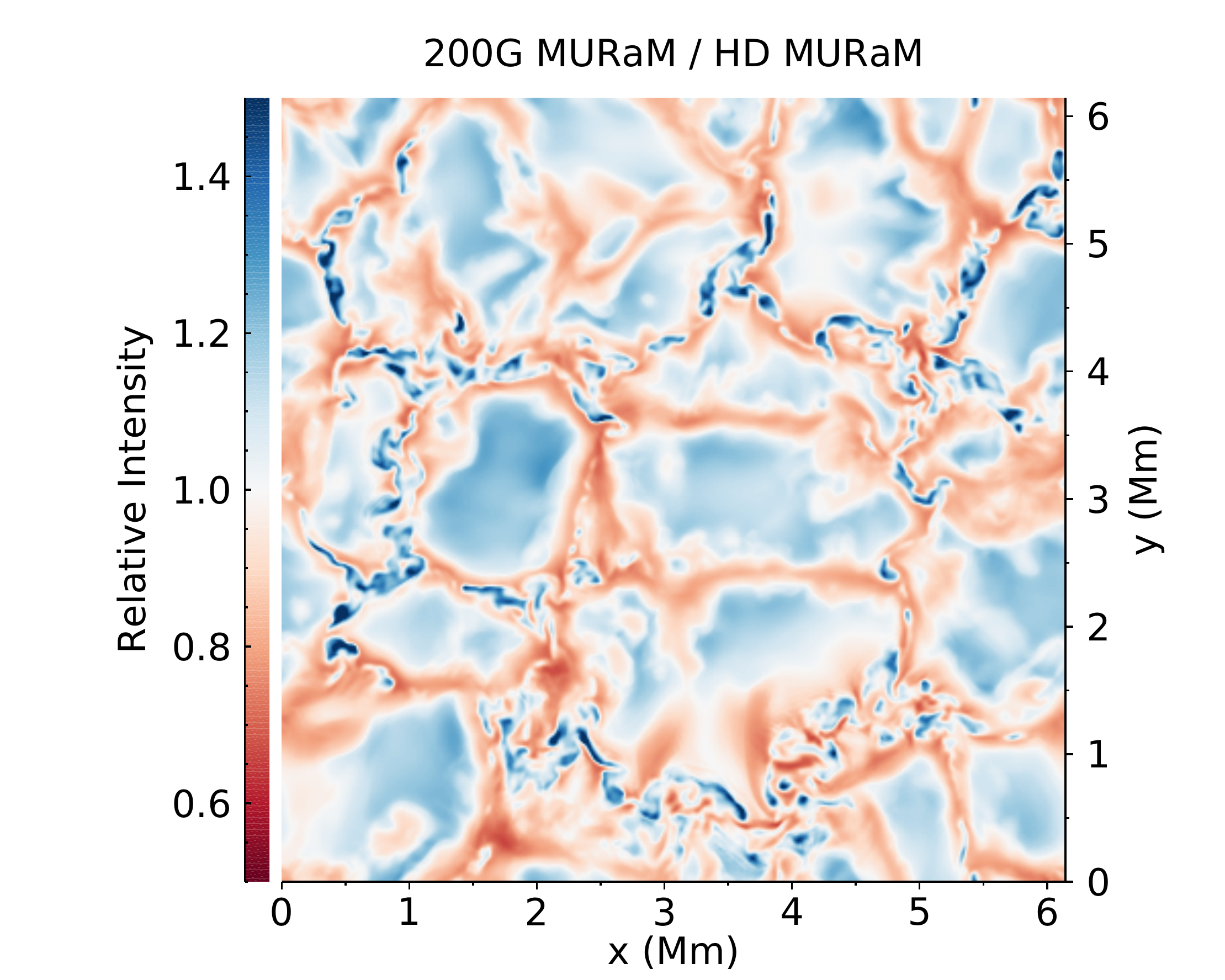}\\
		\includegraphics[width=0.33\linewidth]{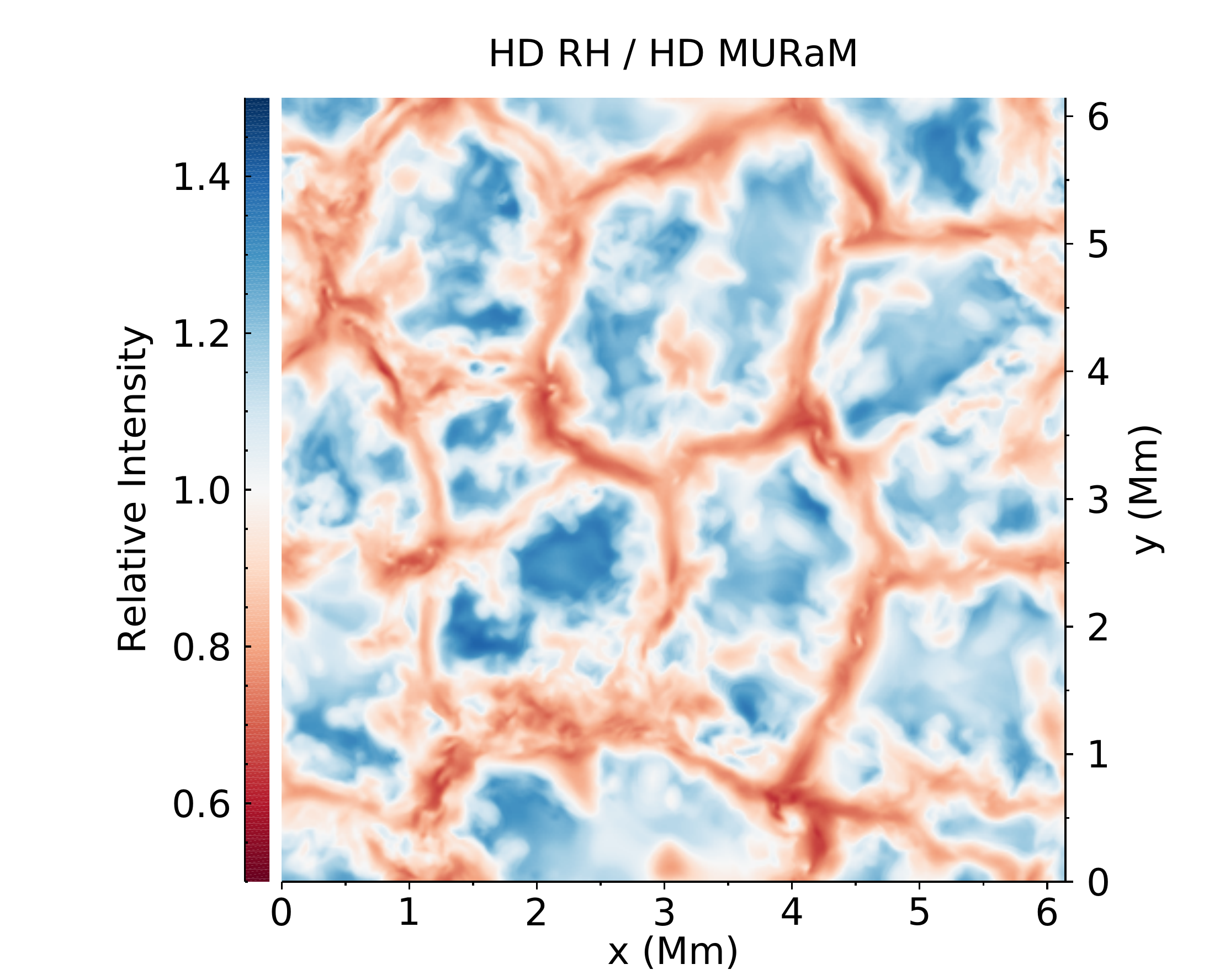}
		\includegraphics[width=0.33\linewidth]{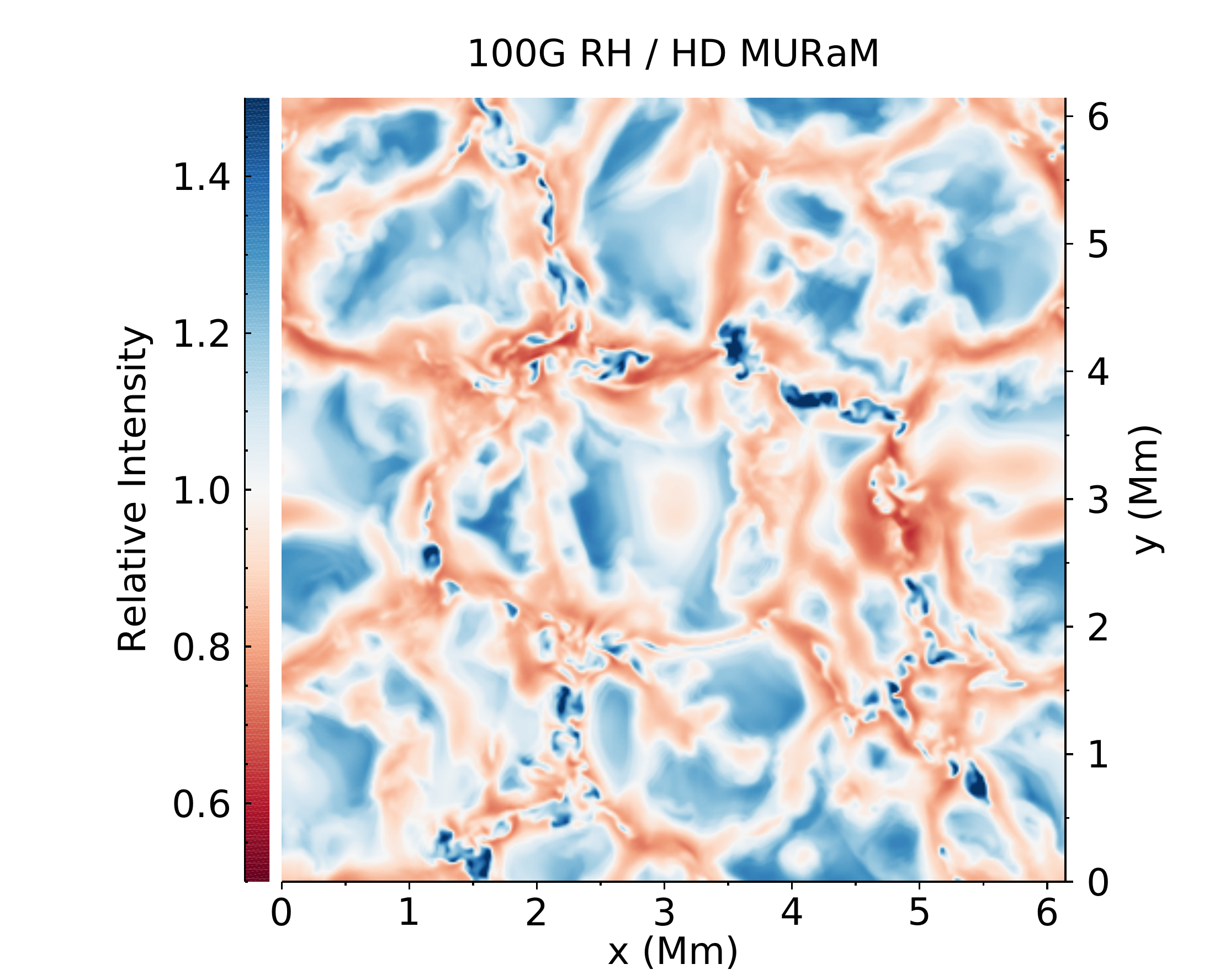}
		\includegraphics[width=0.33\linewidth]{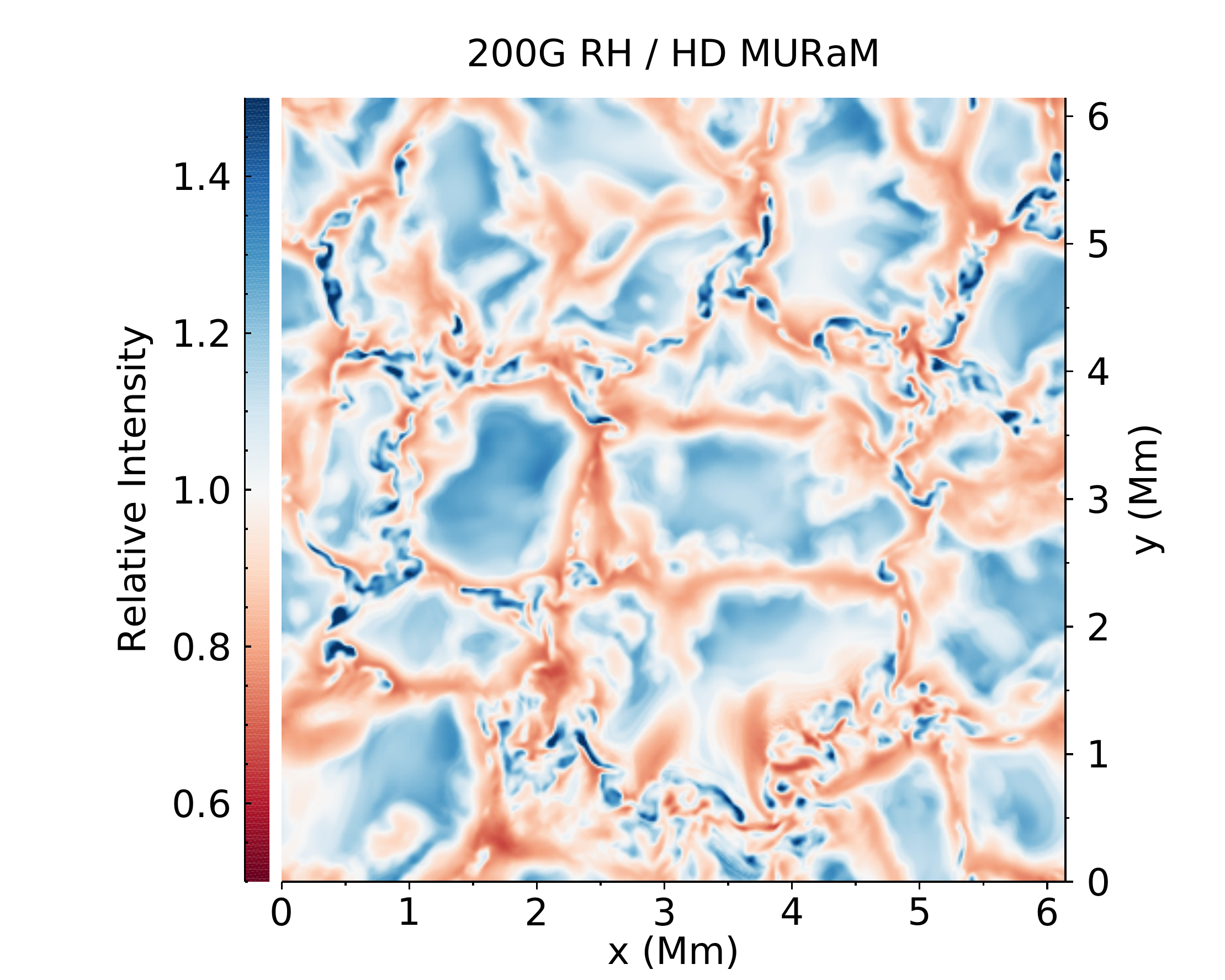}\\
		\includegraphics[width=0.33\linewidth]{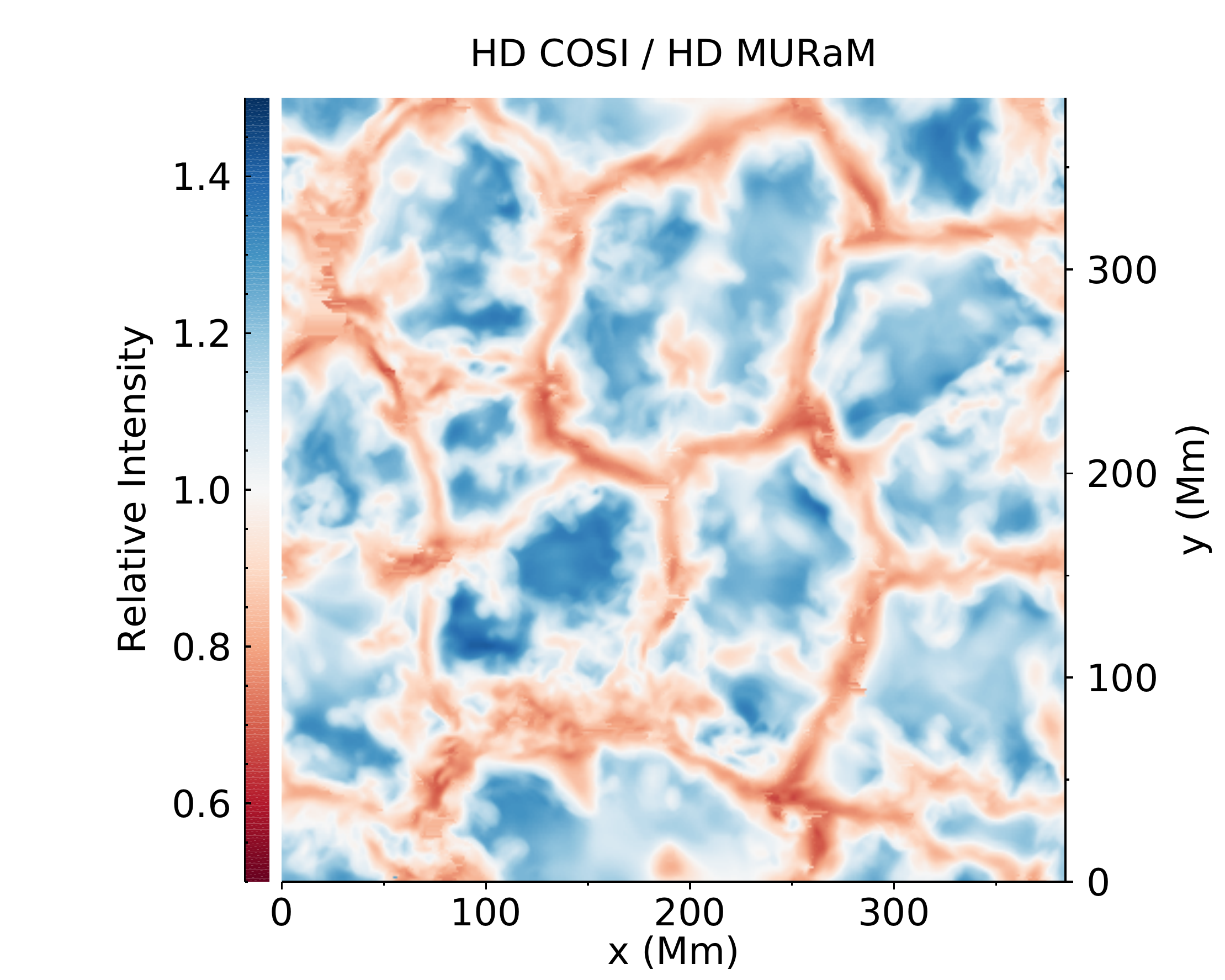}
		\includegraphics[width=0.33\linewidth]{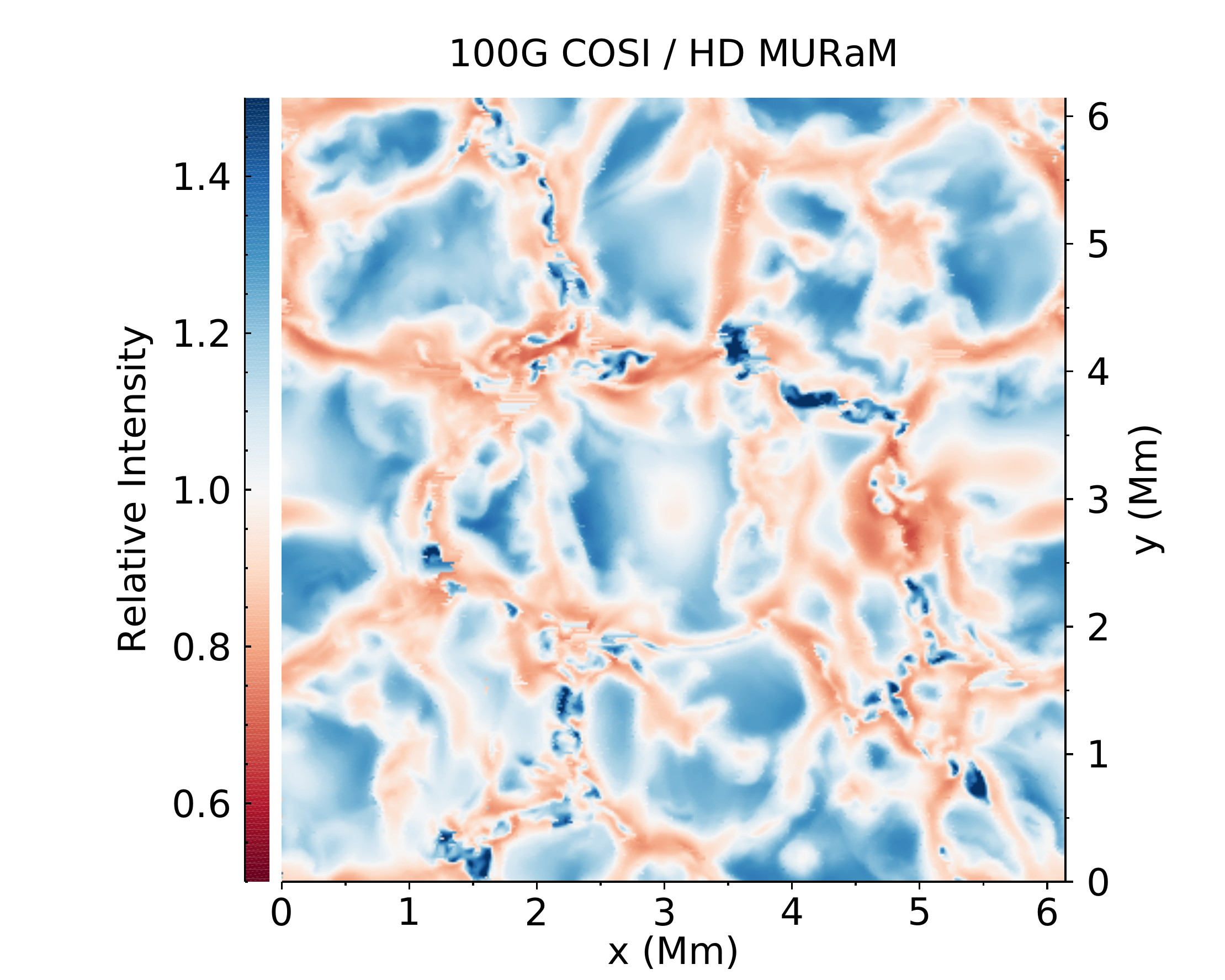}
		\includegraphics[width=0.33\linewidth]{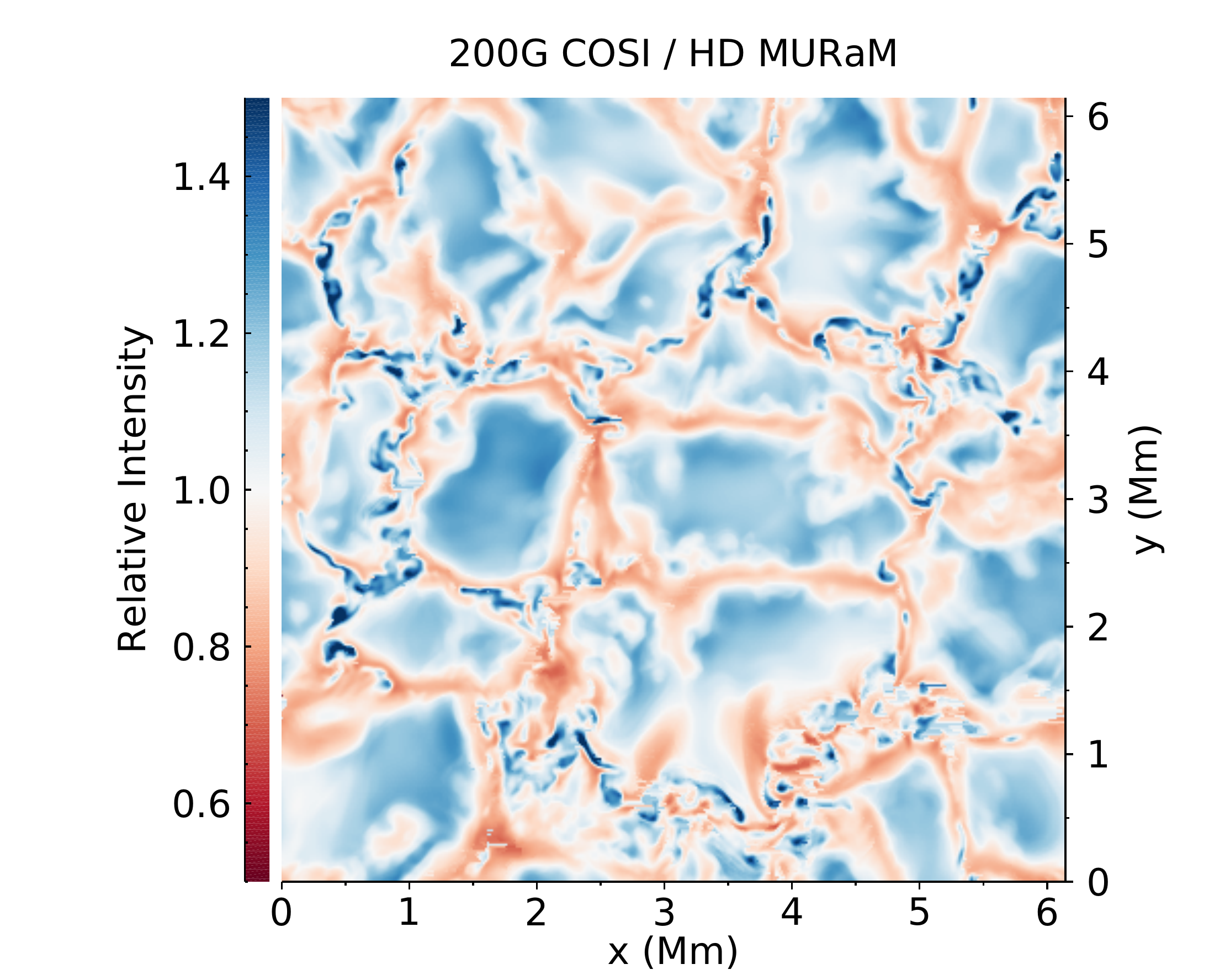}\\
		\caption{\label{fig:contrast_separate1} Relative intensity for the HD, 100 G and 200 G snapshots calculated from the MURaM radiation scheme (top row), the RH code (middle row) and COSI (bottom row). The intensities of all snapshots were  normalized to the mean of the MURaM HD snapshot, which is 2.71 10${^7}$ erg s$^{-1}$ cm$^{-2}$ nm$^{-1}$ sr$^{-1}$. For better visibility of the low-medium contrast features, the color scale only covers the range between 0.5 and above 1.5 and is kept constant outside that range.} 
	\end{center}
\end{figure*}
\begin{figure*}[th!]
	\begin{center}
		\includegraphics[width=0.33\linewidth]{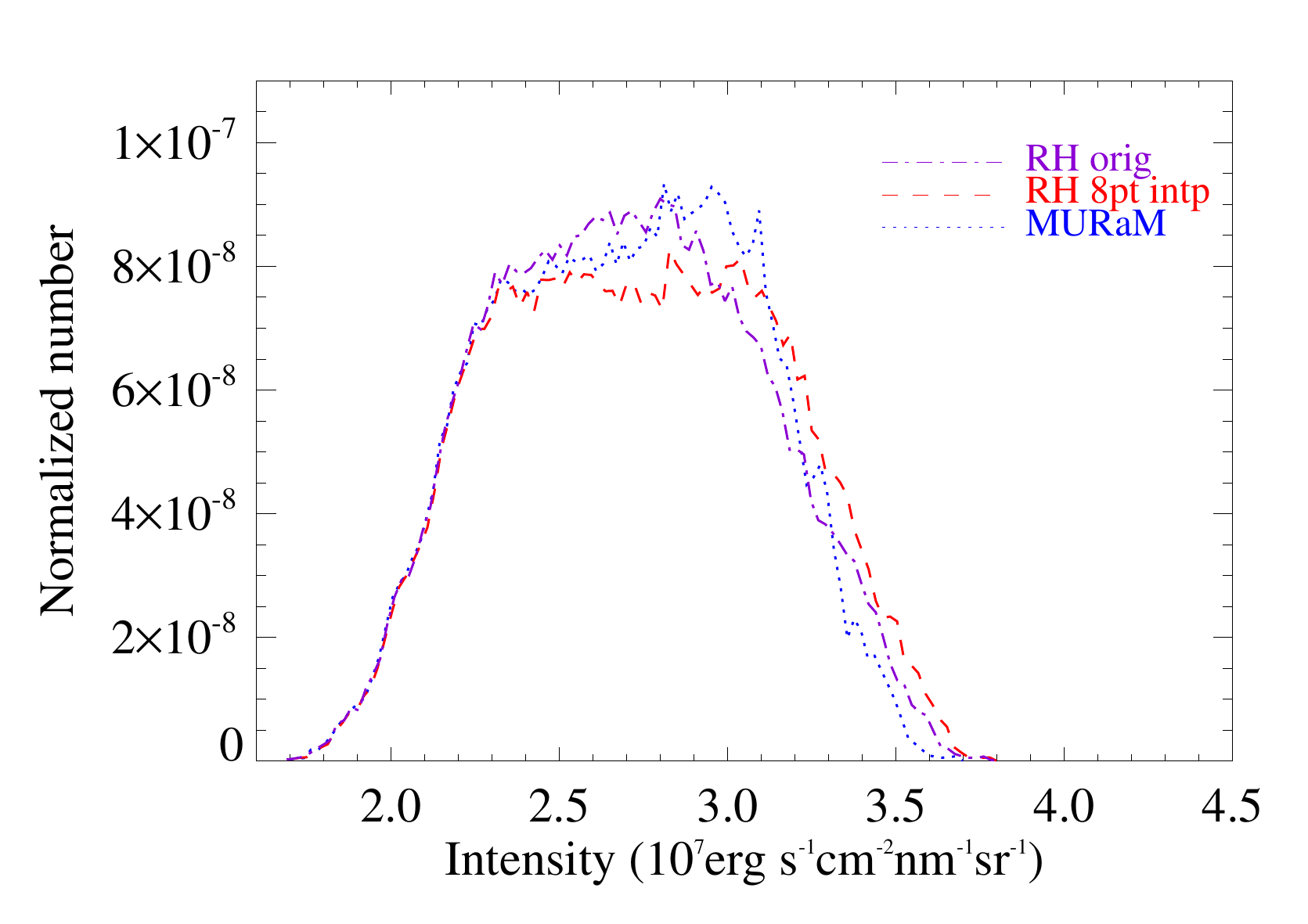}
		\includegraphics[width=0.33\linewidth]{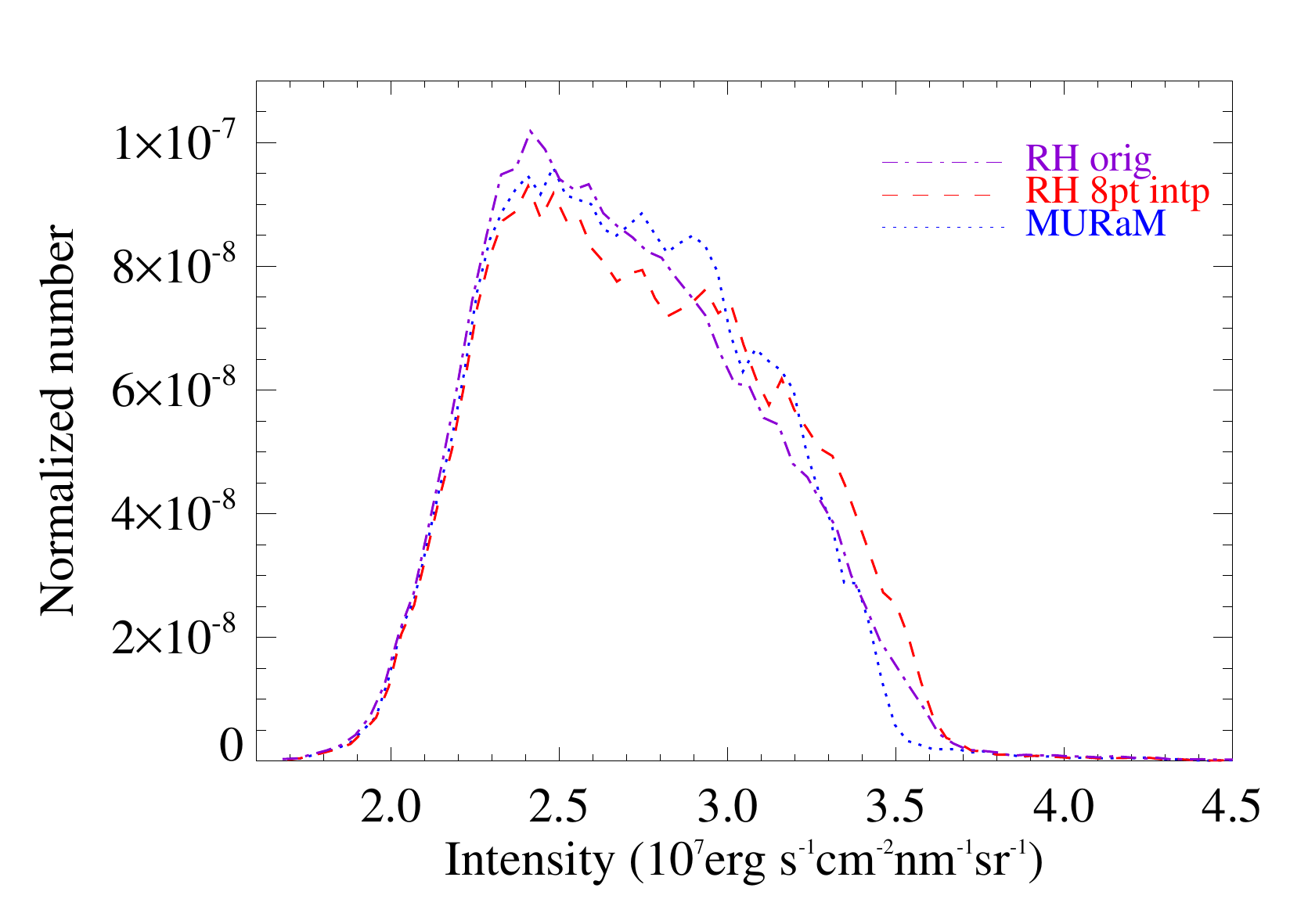}
		\includegraphics[width=0.33\linewidth]{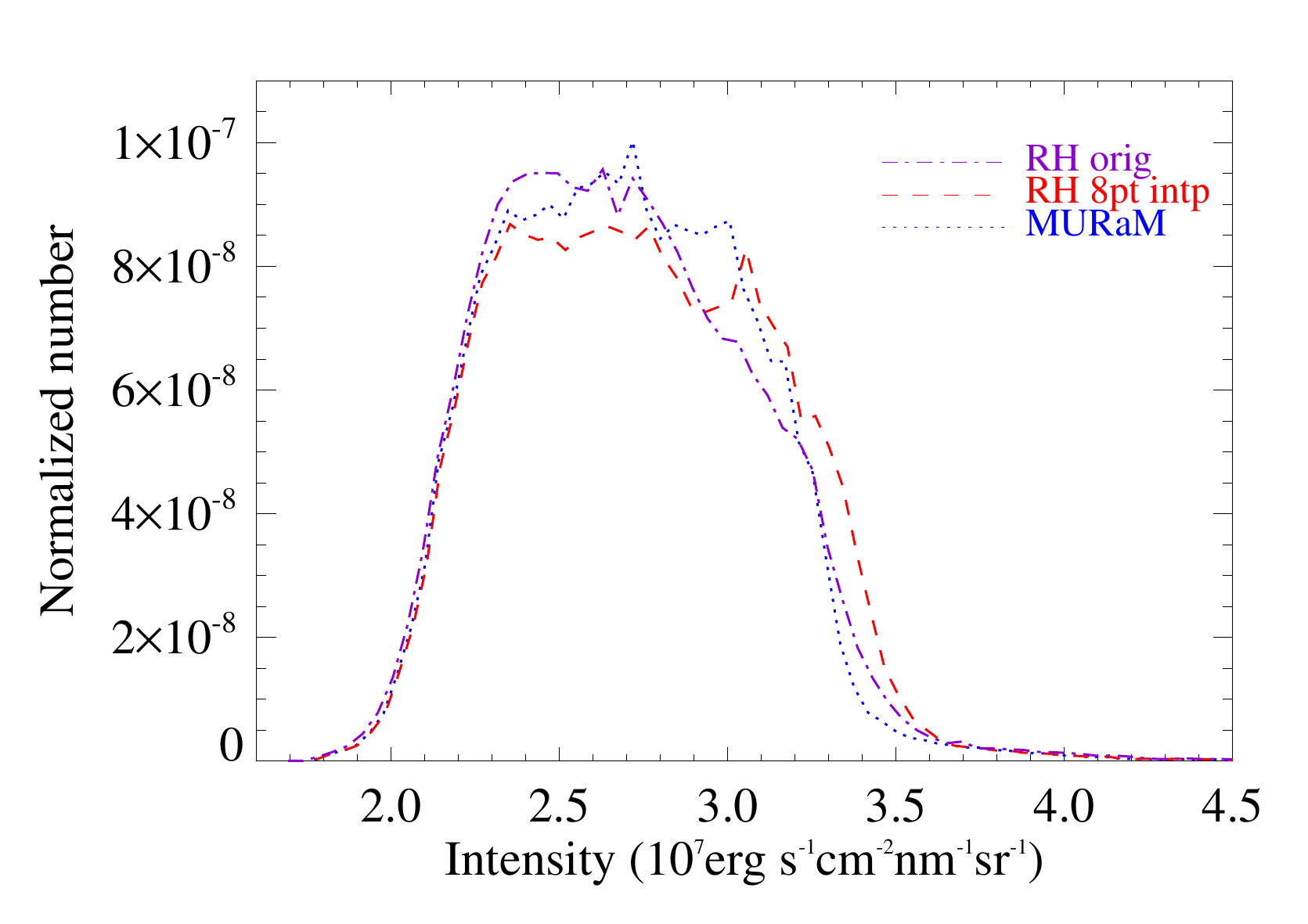} \\
		\includegraphics[width=0.33\linewidth]{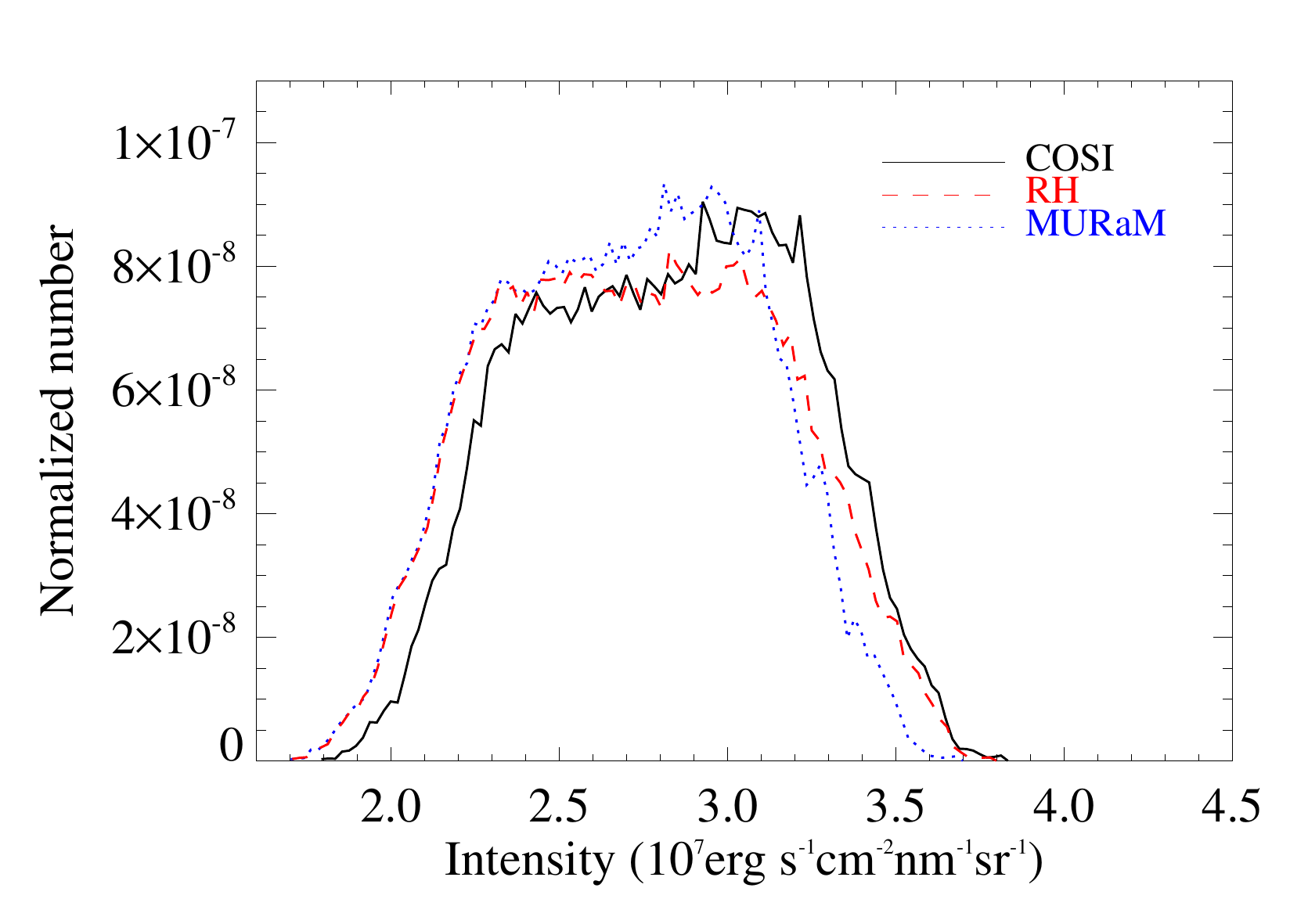}
		\includegraphics[width=0.33\linewidth]{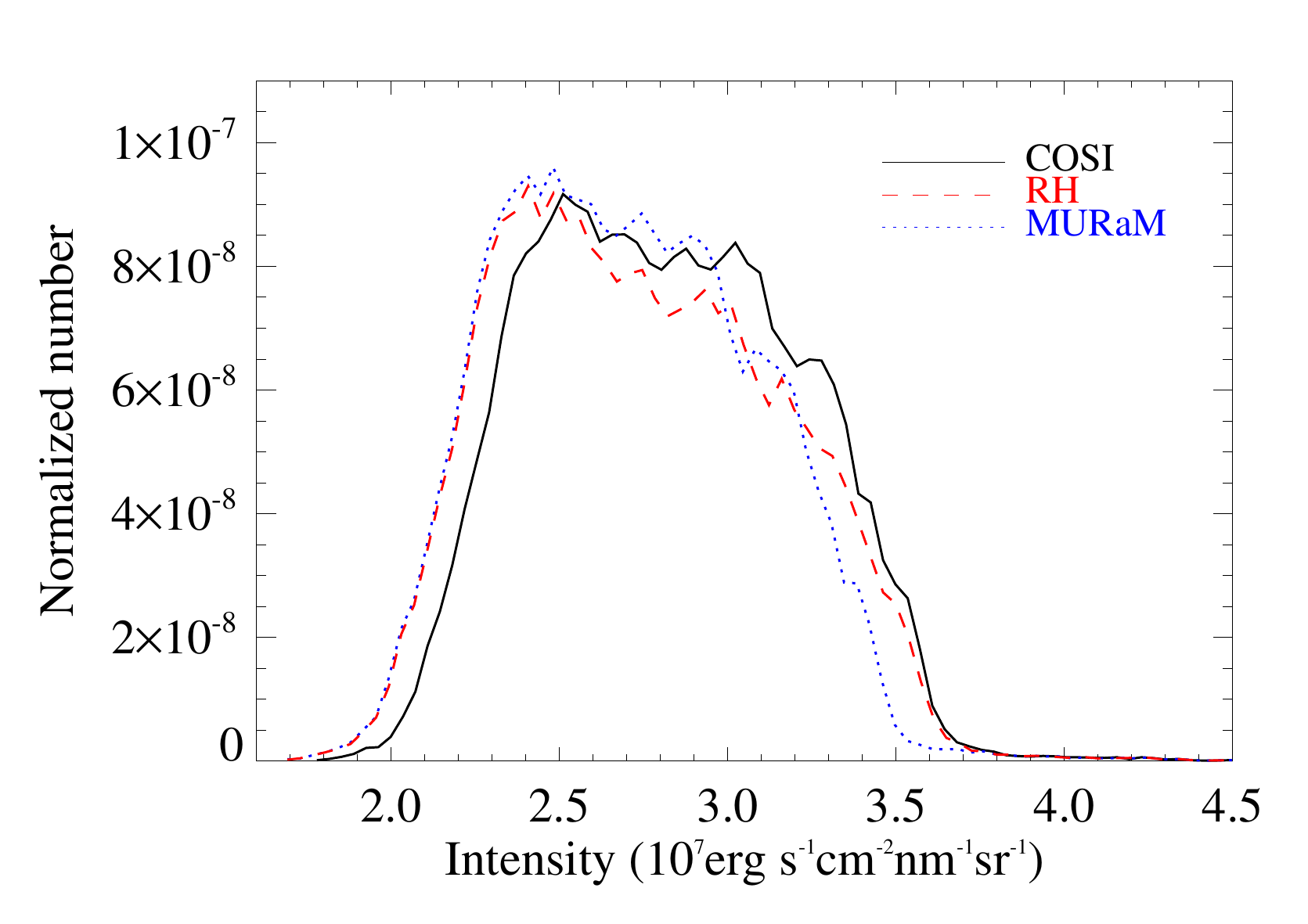}
		\includegraphics[width=0.33\linewidth]{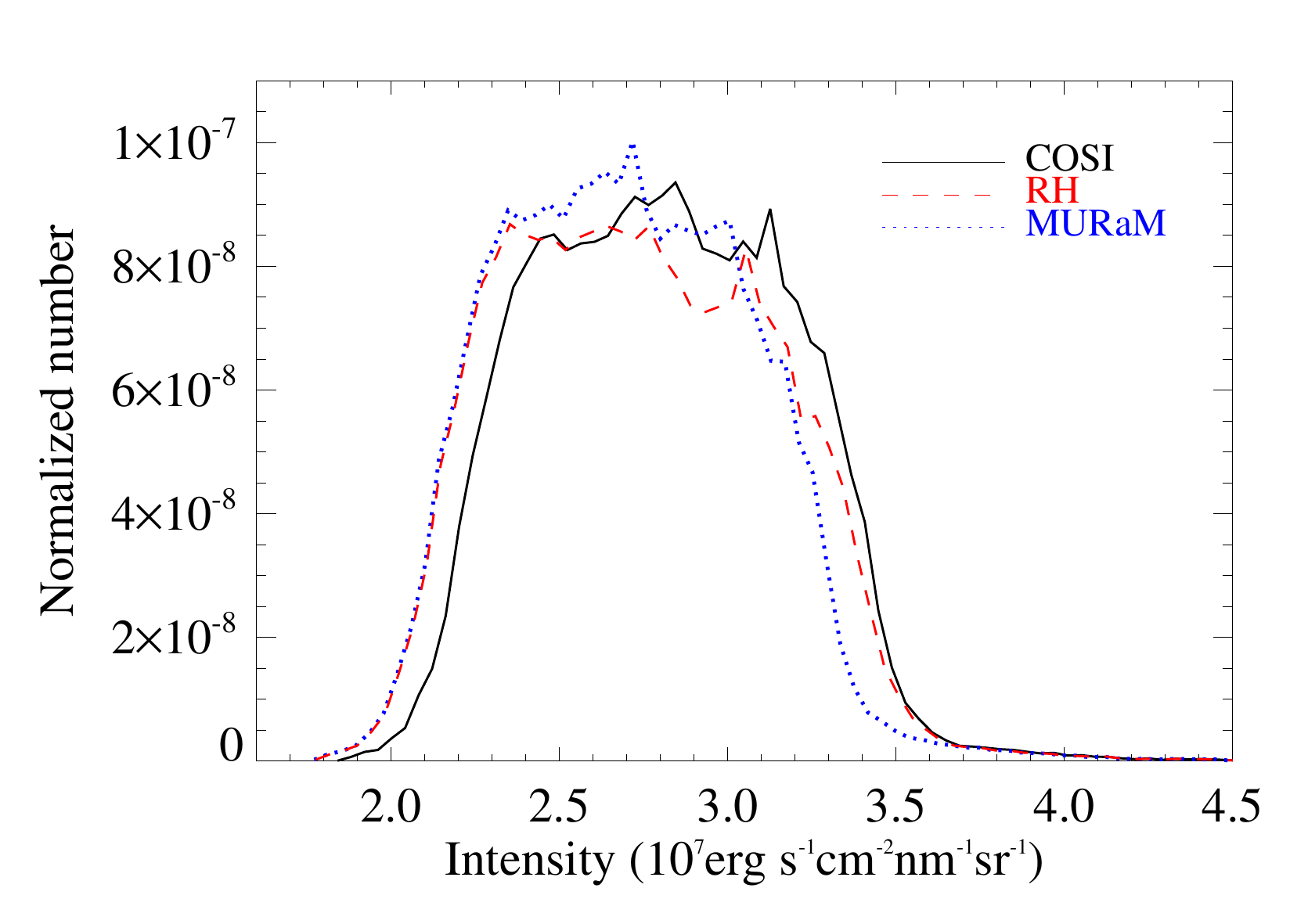} 
		\caption{\label{fig:distribution} Top panel: intensity distribution function for the full $384\times384$ snapshot calculated with the RH code using the original grid in the MHD simulation ({purple dot-dashed} lines) and the 8-point interpolated MHD grid (red dashed lines) and the MURaM calculation (blue lines). Bottom Panel: intensity distribution function for the HD (left panel), 100 G (middle panel) and 200 G (right panel) snapshots using the COSI (black lines), MURaM solver (blue lines), and RH (red lines) codes.} 
	\end{center}
\end{figure*}
\section{Spectral synthesis codes}{\label{sec:codes}}
For the spectral synthesis based on the 3D MHD simulations we use three different radiative transfer codes. First, the COde for Solar Irradiance \citep[COSI,][]{Haberreiter2008a,Haberreiter2008b,shapiro2010,Criscuoli2020} allows us to calculate the atomic level populations and the emergent intensity taking into account non-local thermodynamic equilibrium (non-LTE), which is a key element for the calculation of the UV spectral range. {The atomic data used in COSI are explained in detail in \cite{shapiro2010} and the numerical scheme of the radiative transfer goes back to \cite{HamannSchmutz1987} and \cite{Hubeny1981}. The scheme was first applied to solar studies by \cite{Haberreiter2008a,Haberreiter2008b}.} {So far}, COSI {calculations have been based on} vertical atmosphere structures, such as the 1D atmosphere structures by \cite{fontenla1999}. {In this work}, we use for the first time 3D MHD simulations as input to COSI.

\begin{figure*}[th!]
	\begin{center}
		\includegraphics[width=0.33\linewidth]{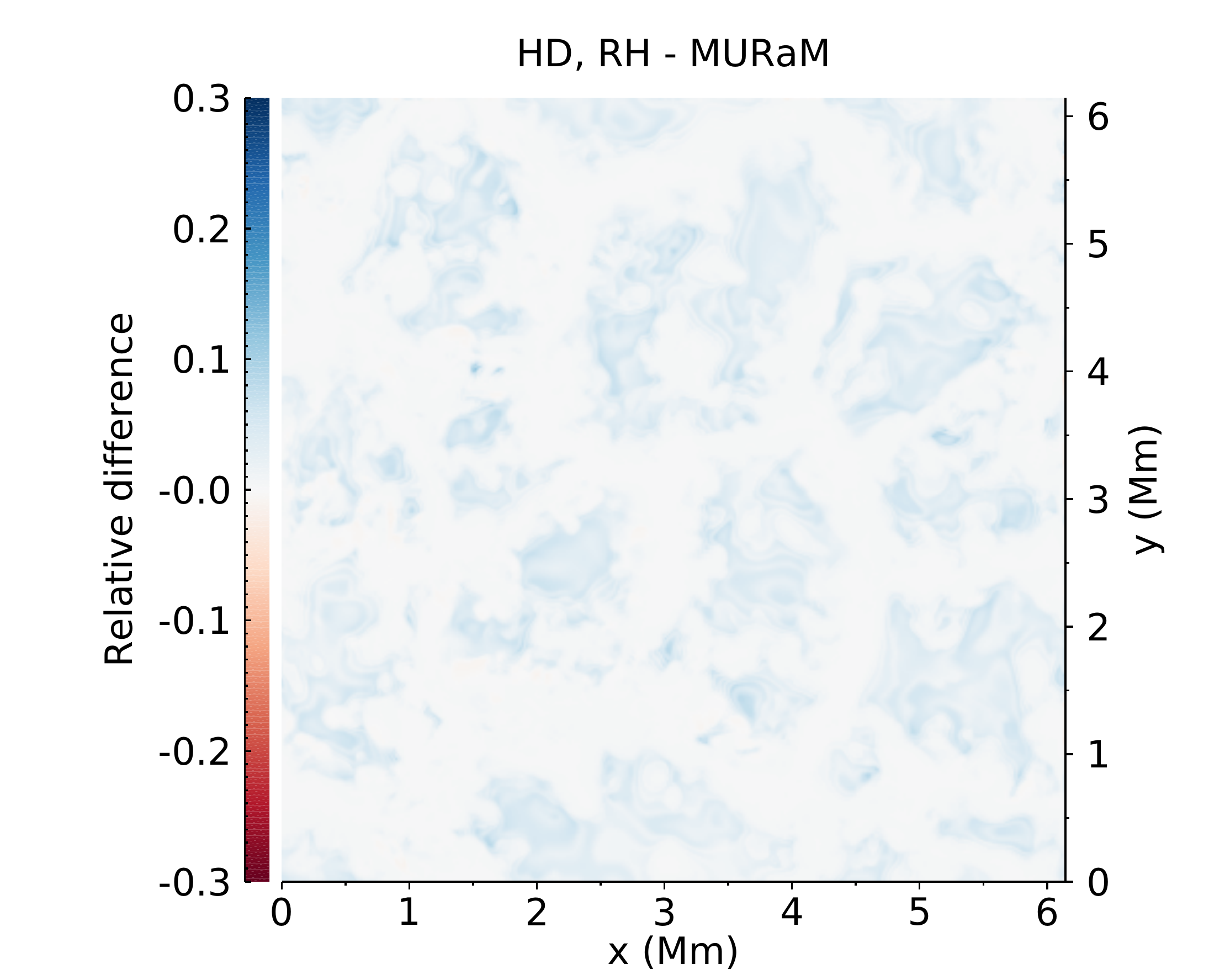}
		\includegraphics[width=0.33\linewidth]{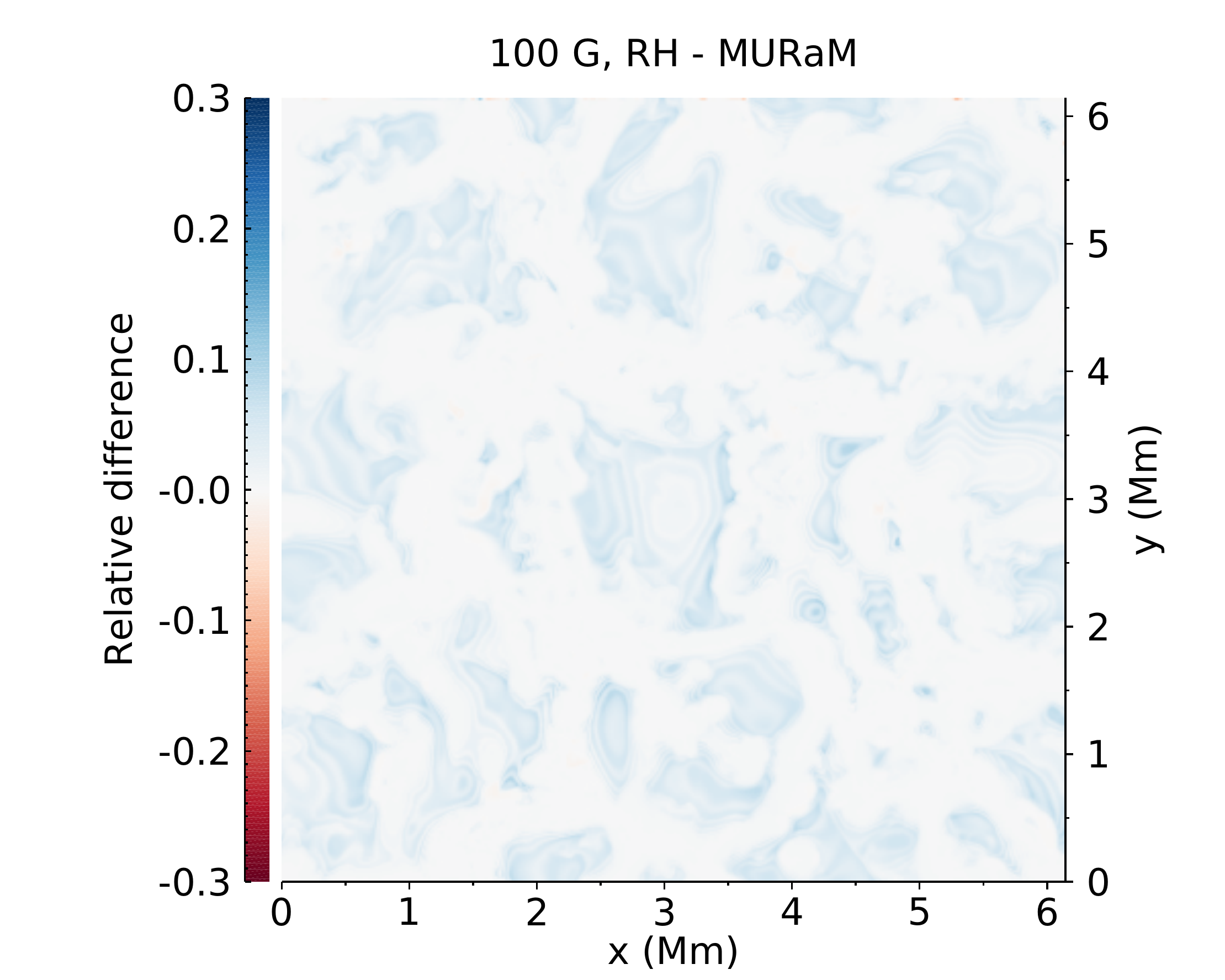}
		\includegraphics[width=0.33\linewidth]{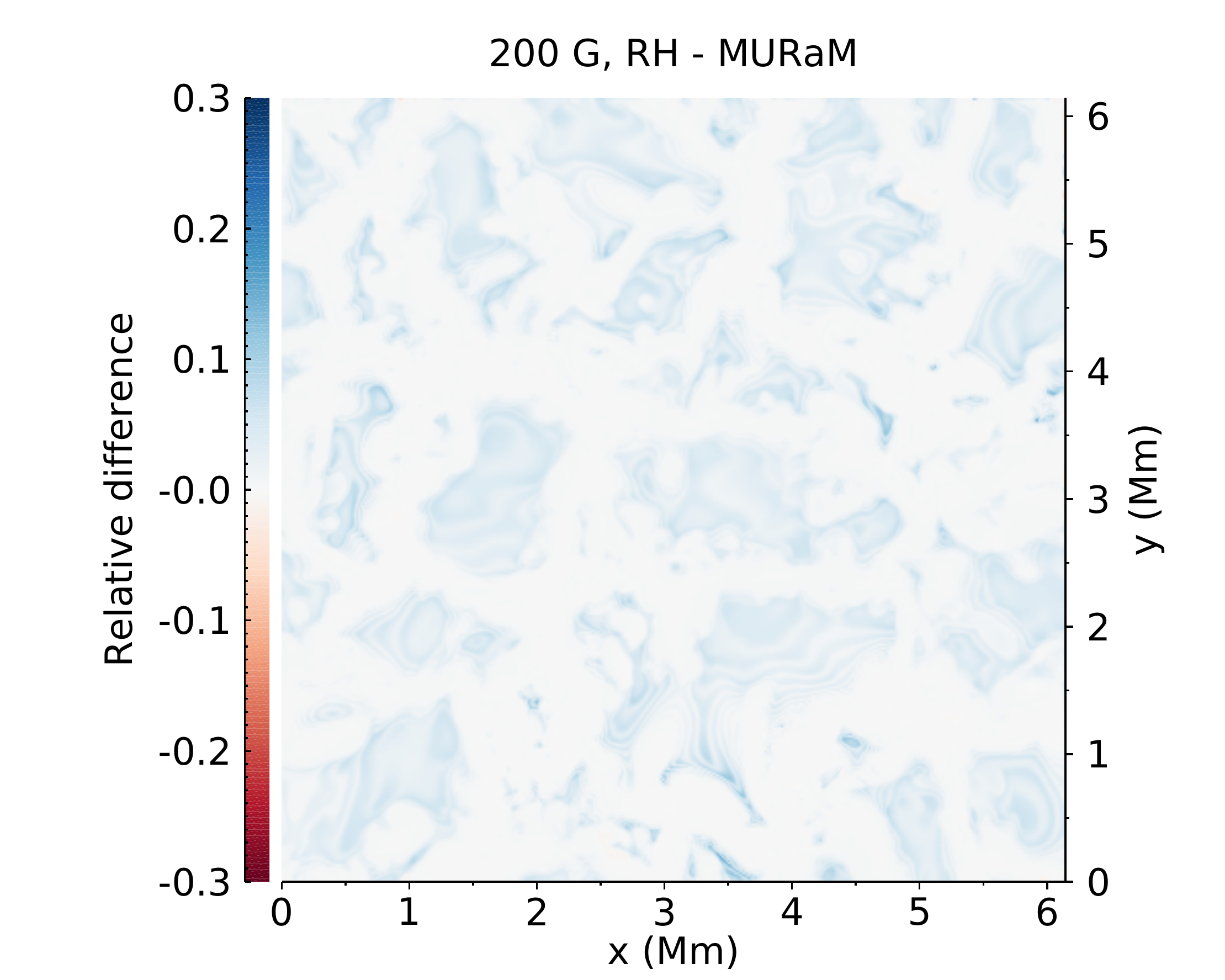}\\
		\includegraphics[width=0.33\linewidth]{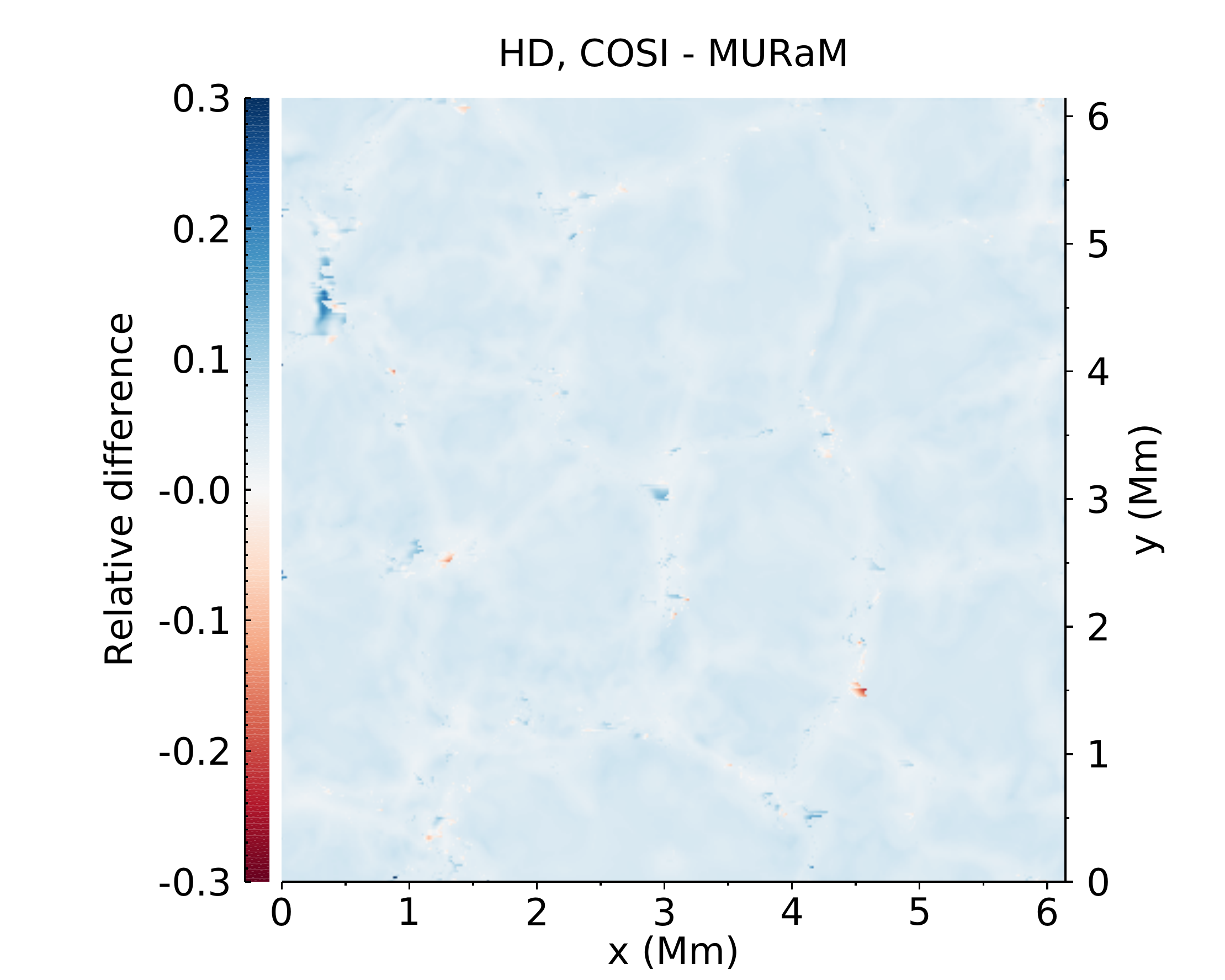}
		\includegraphics[width=0.33\linewidth]{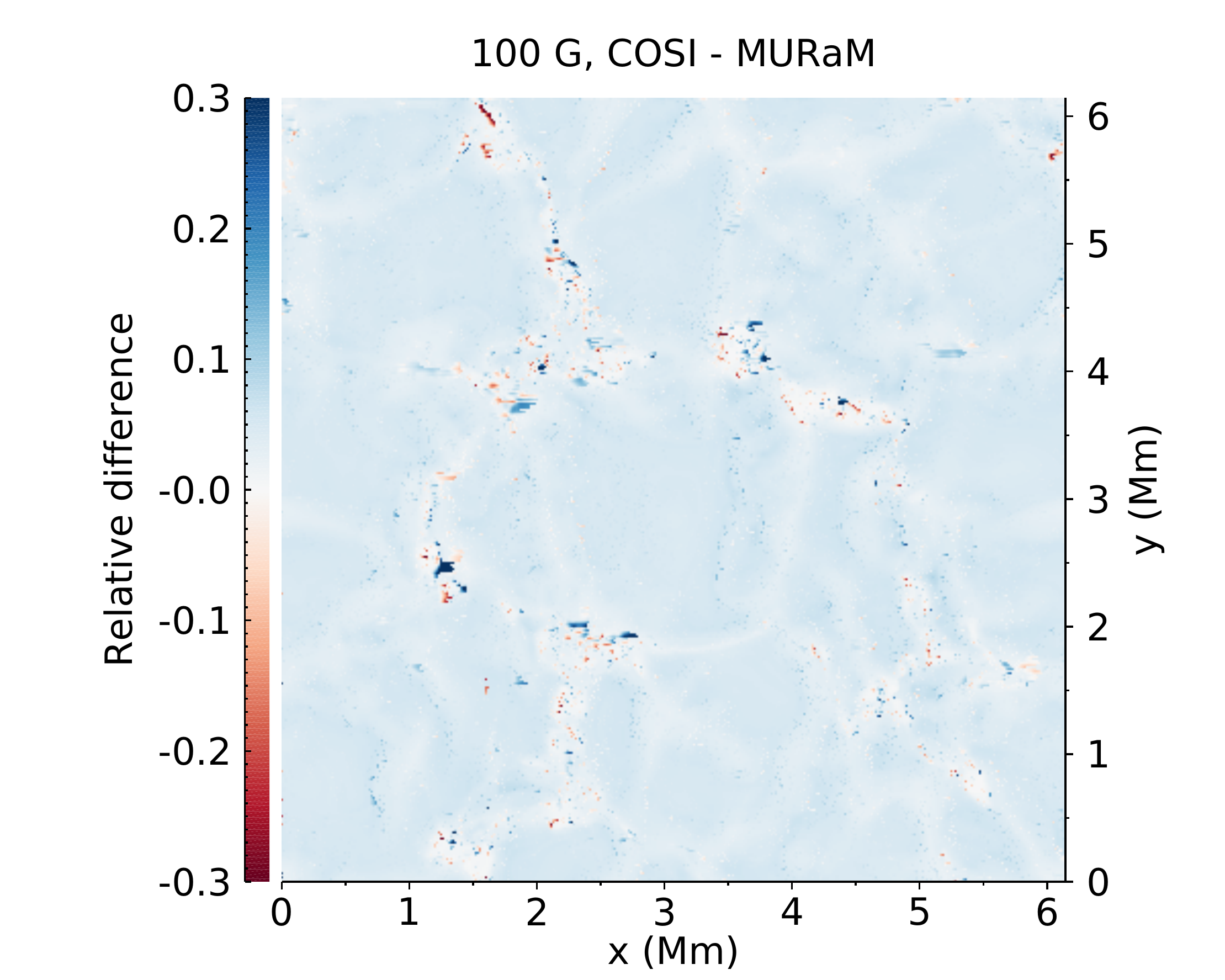}
		\includegraphics[width=0.33\linewidth]{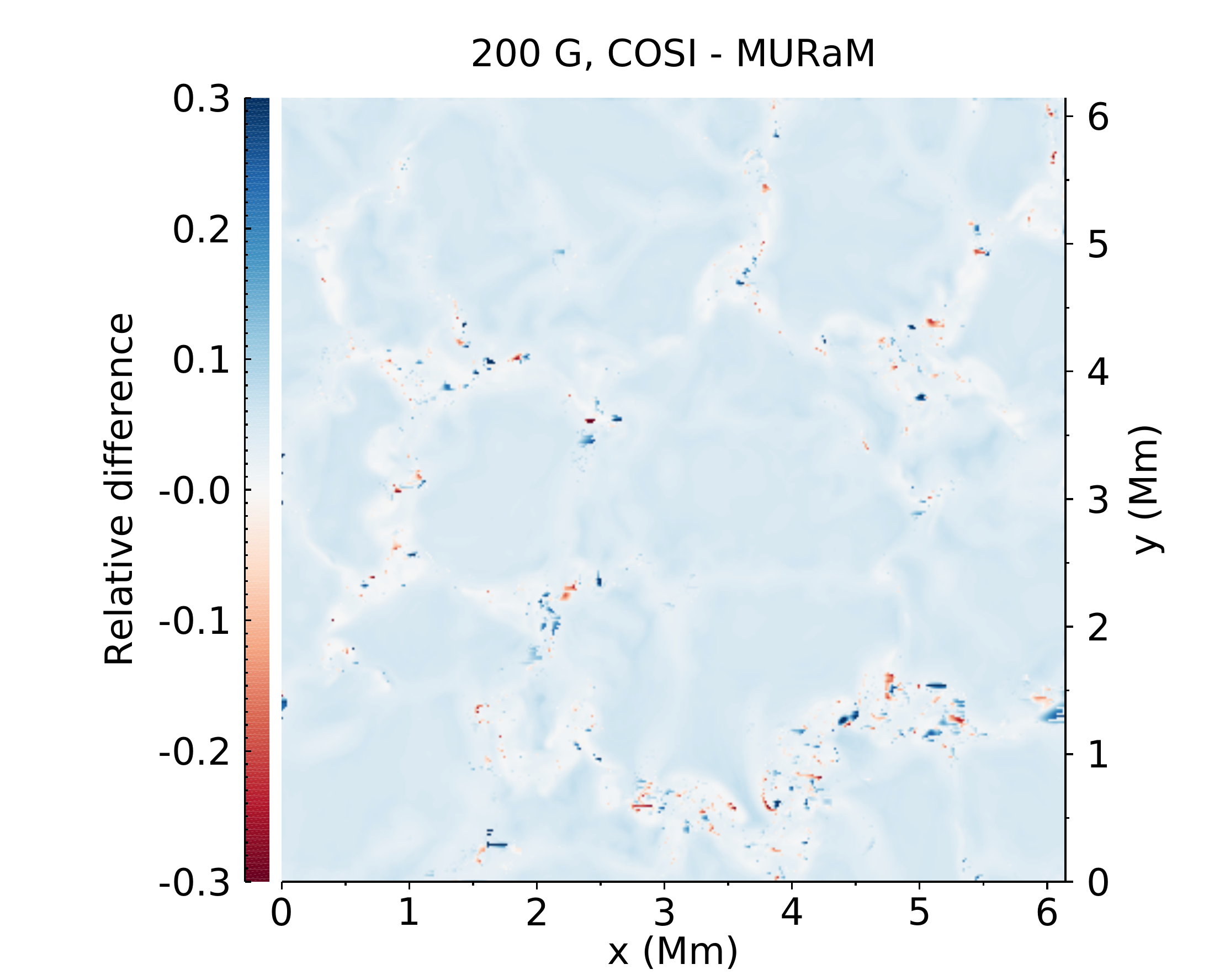}\\
		\caption{\label{fig:diff} Top panels: relative pixel-to-pixel difference, (RH$_i$ - MURaM$_i$)/<MURaM,HD>, for the HD, 100 G and 200 G snapshots; Bottom panels: same as top panels but for (COSI$_i$ - MURaM$_i$)/<MURaM,HD>. For better visibility of the low-medium contrast features, the color scale only covers the given ranges between in the top and bottom panels and is kept constant outside of that range.}
	\end{center}
\end{figure*}		
\begin{figure*}[th!]
	\begin{center}		
		\includegraphics[width=0.33\linewidth]{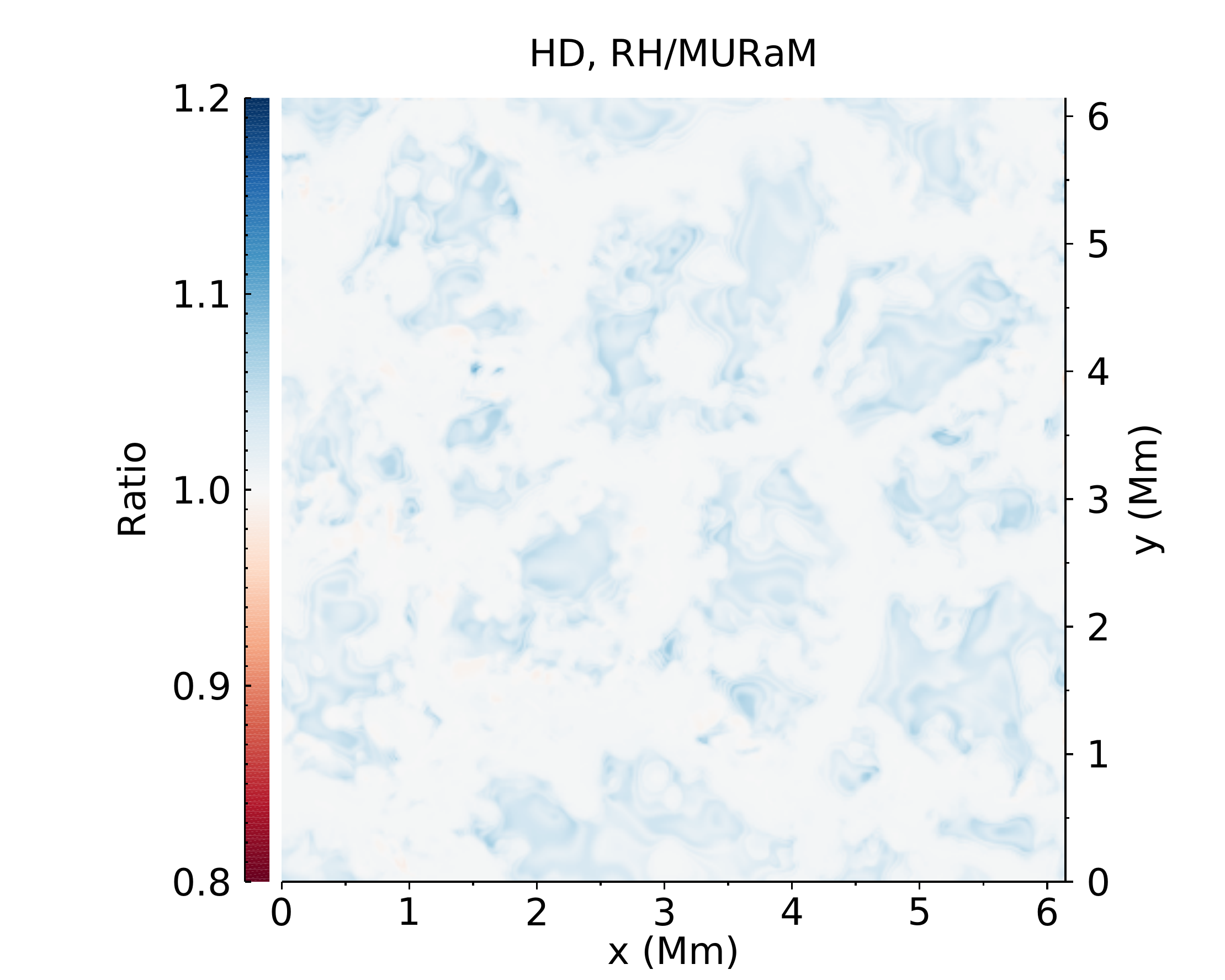}
		\includegraphics[width=0.33\linewidth]{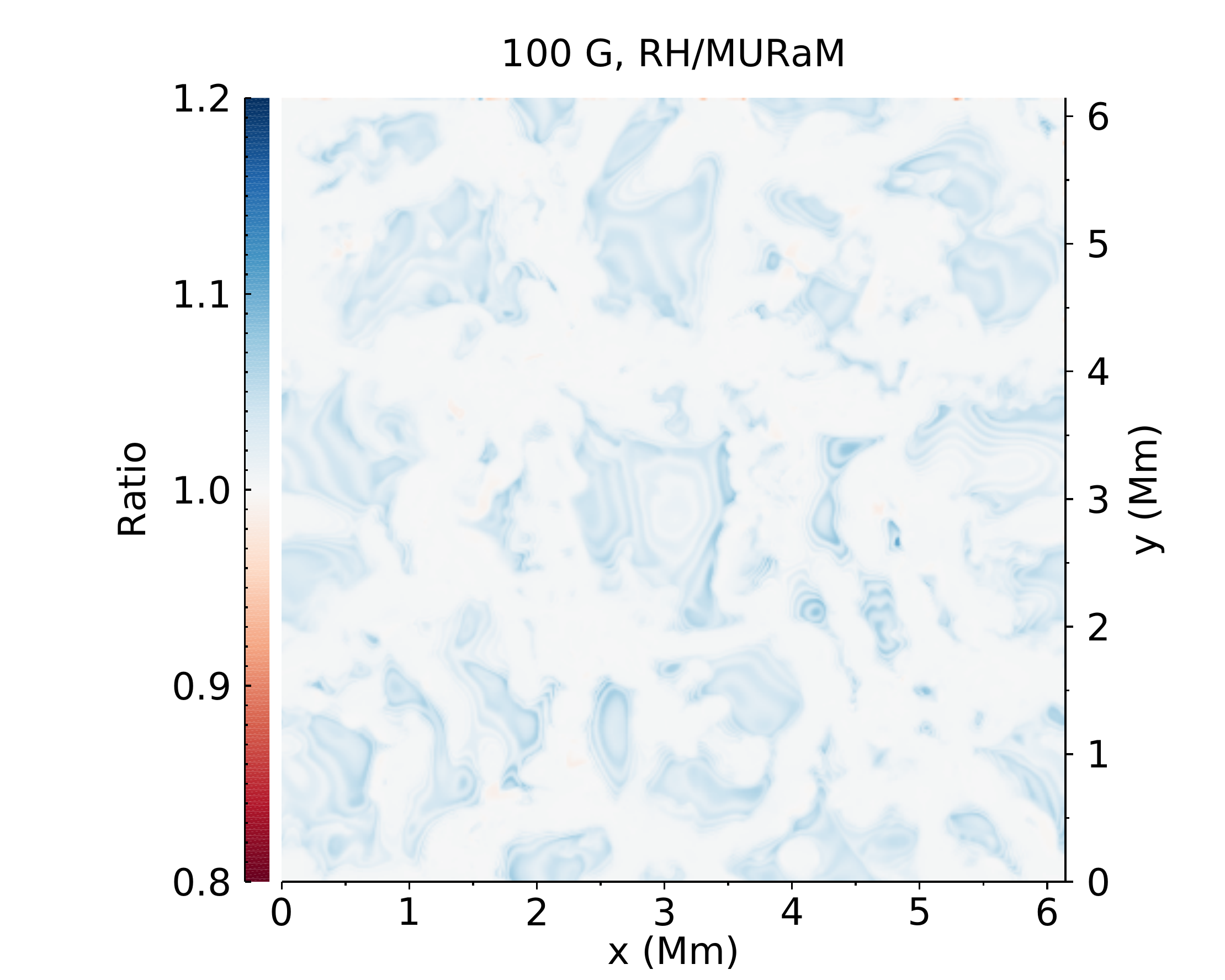}
		\includegraphics[width=0.33\linewidth]{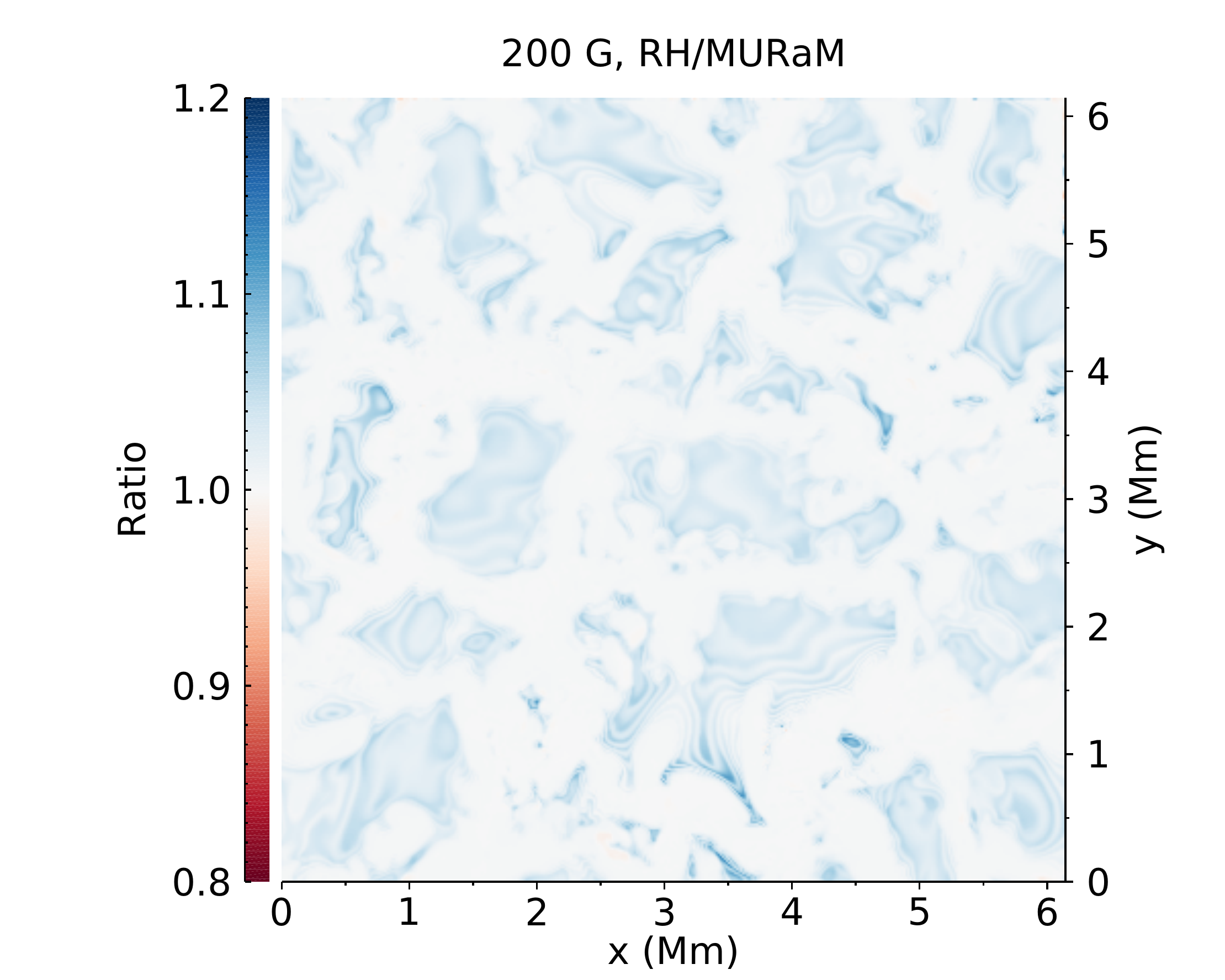}\\

		\includegraphics[width=0.33\linewidth]{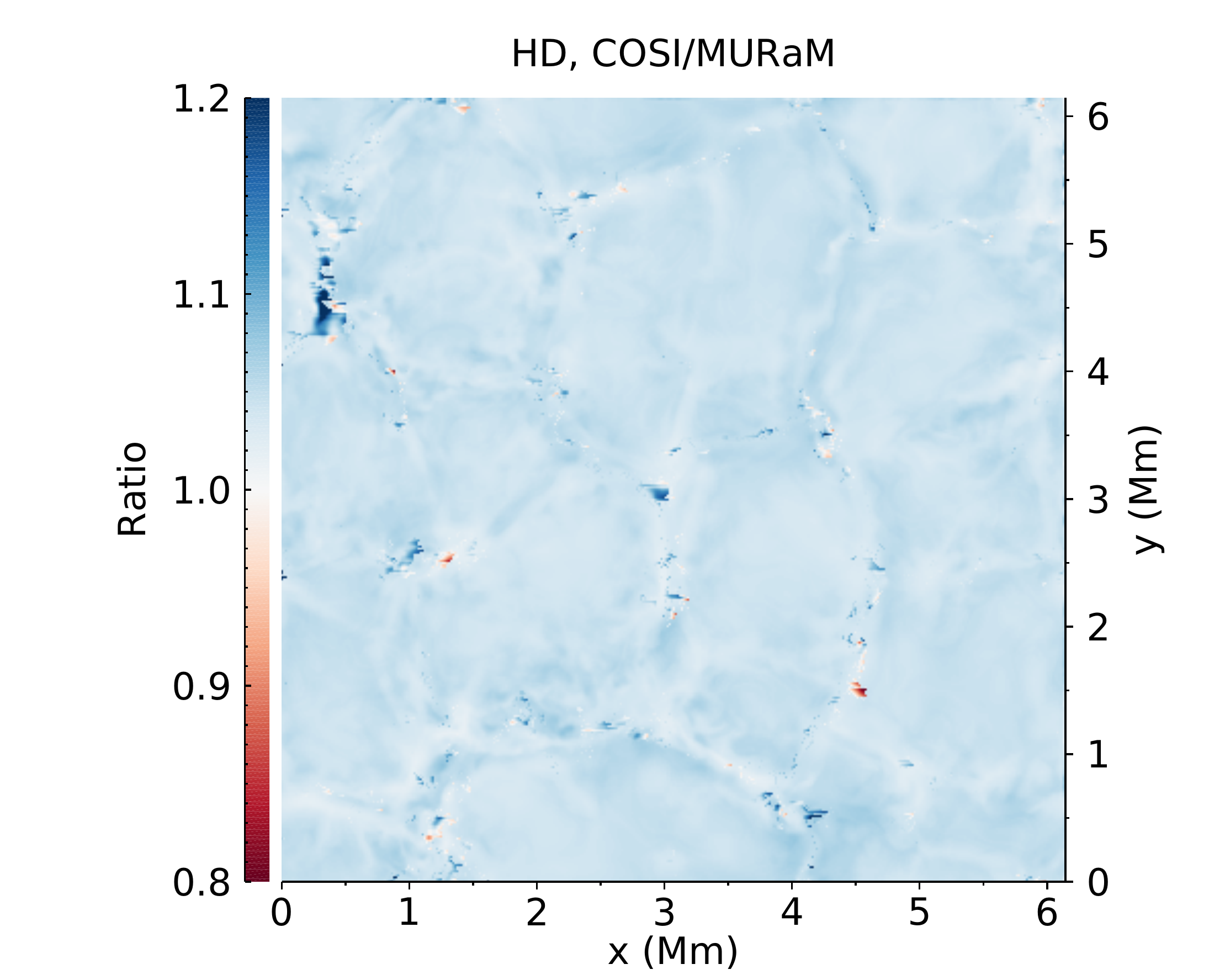}
		\includegraphics[width=0.33\linewidth]{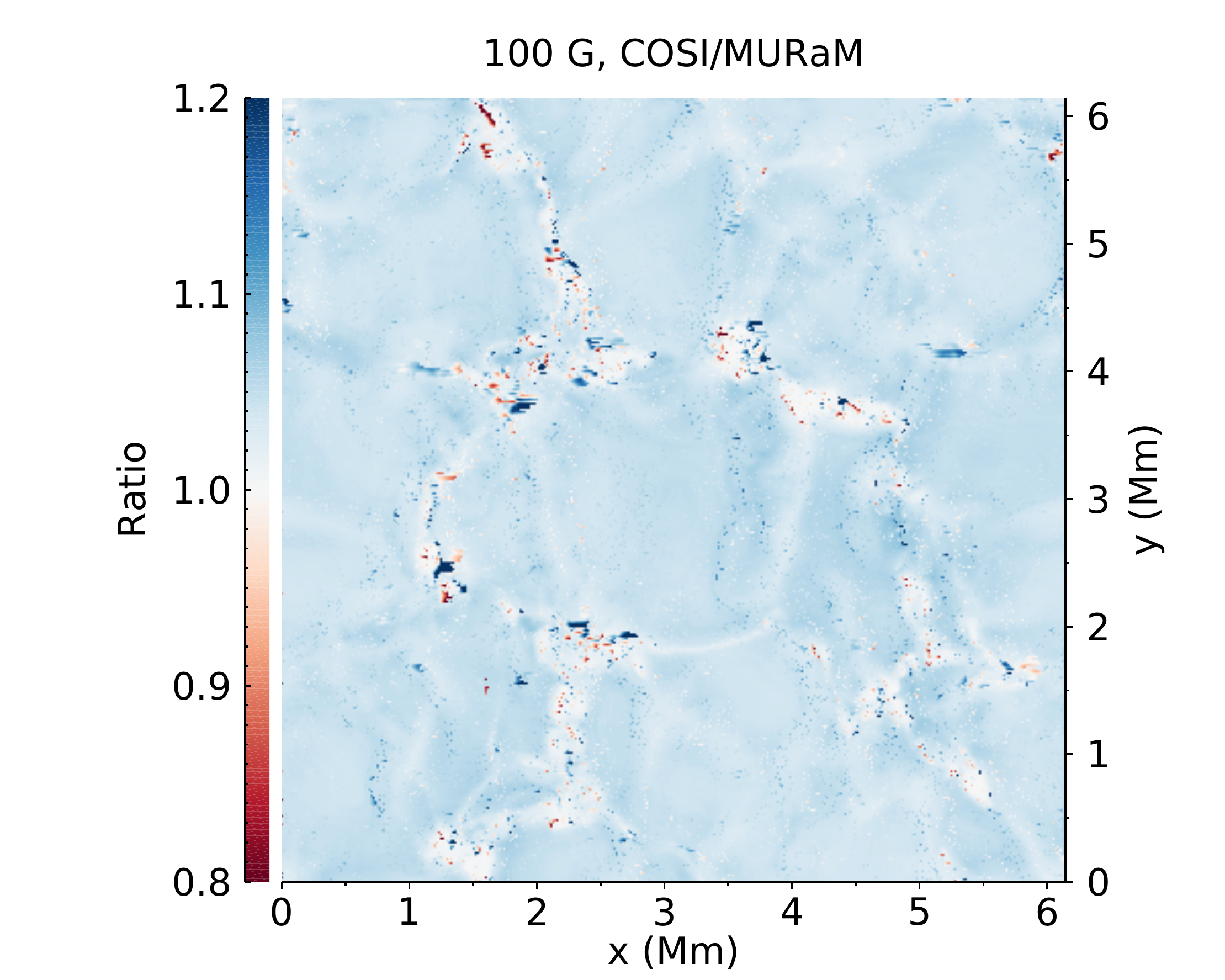}
		\includegraphics[width=0.33\linewidth]{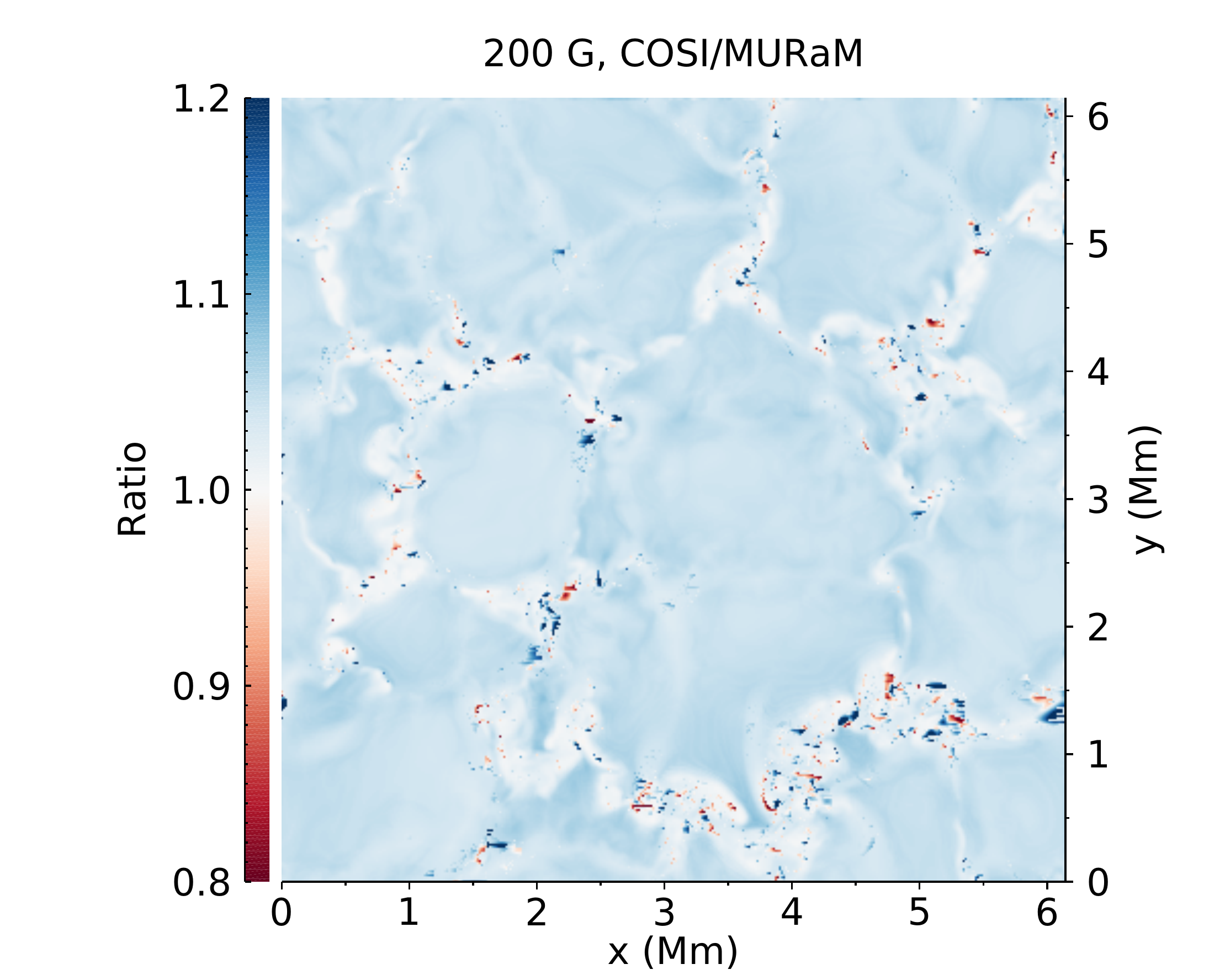}\\		
		\caption{\label{fig:ratio} Top panels: ratio of the intensities, RH$_i$/MURaM$_i$, with $i$ referring to the HD, 100 G and 200 G snapshots, respectively; Bottom panels: same as top panel but for COSI$_i$/MURaM$_i$. For better visibility of the low-medium contrast features, the color scale only covers the given ranges between in the top and bottom panels and is kept constant outside of that range.}
	\end{center}
\end{figure*}
\begin{figure*}[th!]
	\begin{center}
		\hspace{-0.cm}\includegraphics[width=0.33\linewidth]{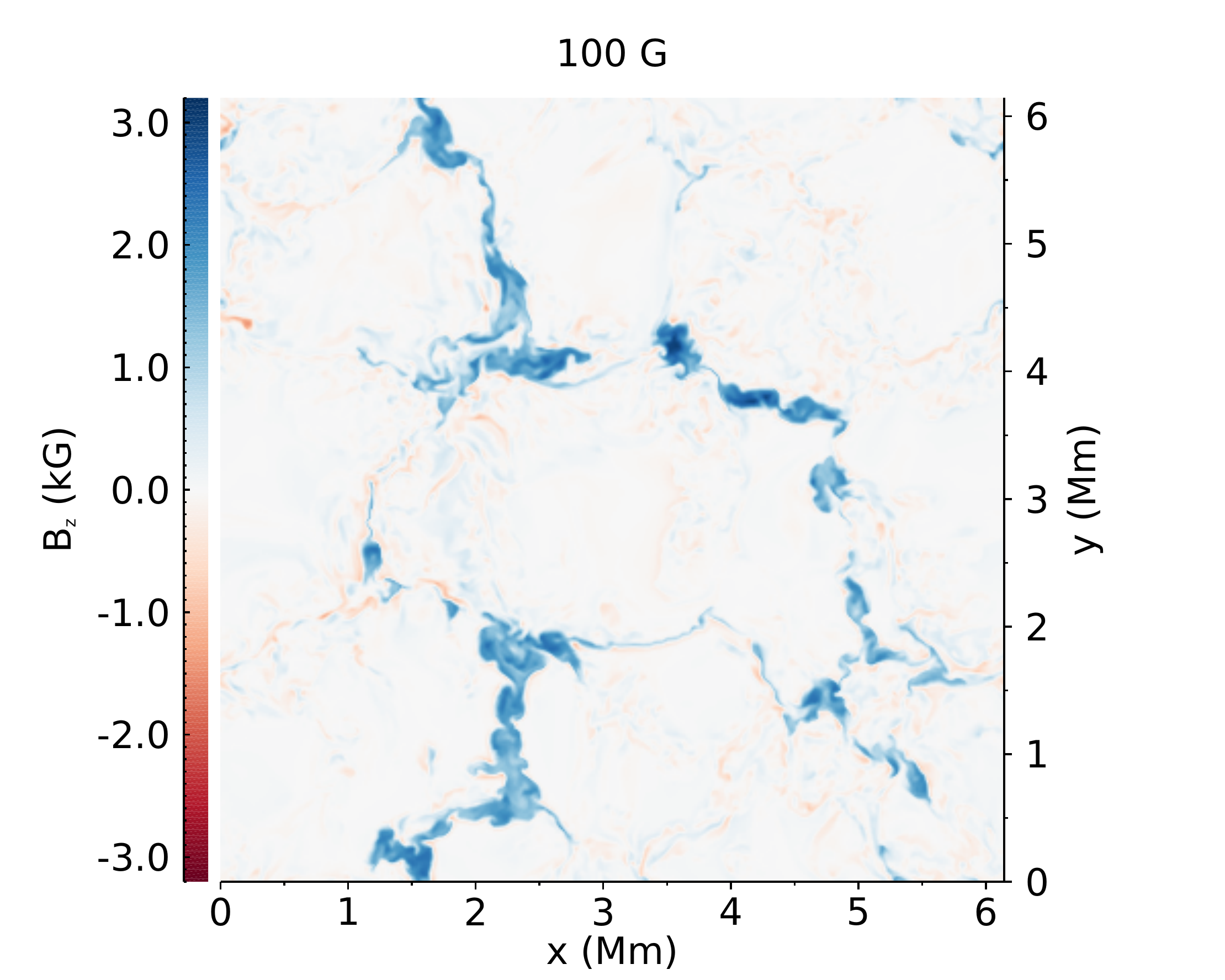}
		\hspace*{-0.cm}\includegraphics[width=0.33\linewidth]{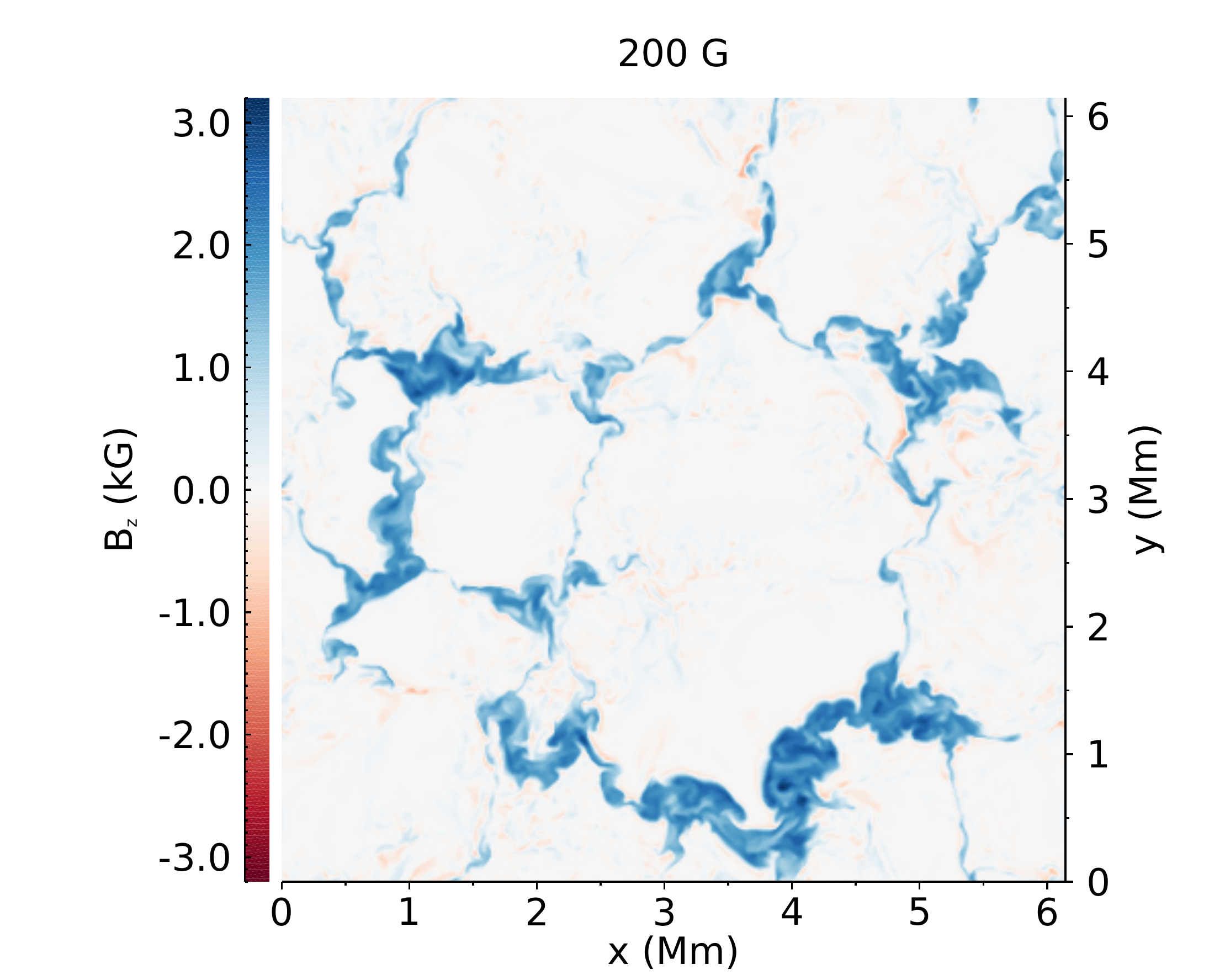}\\
		\hspace*{0.3cm}\includegraphics[width=0.33\linewidth]{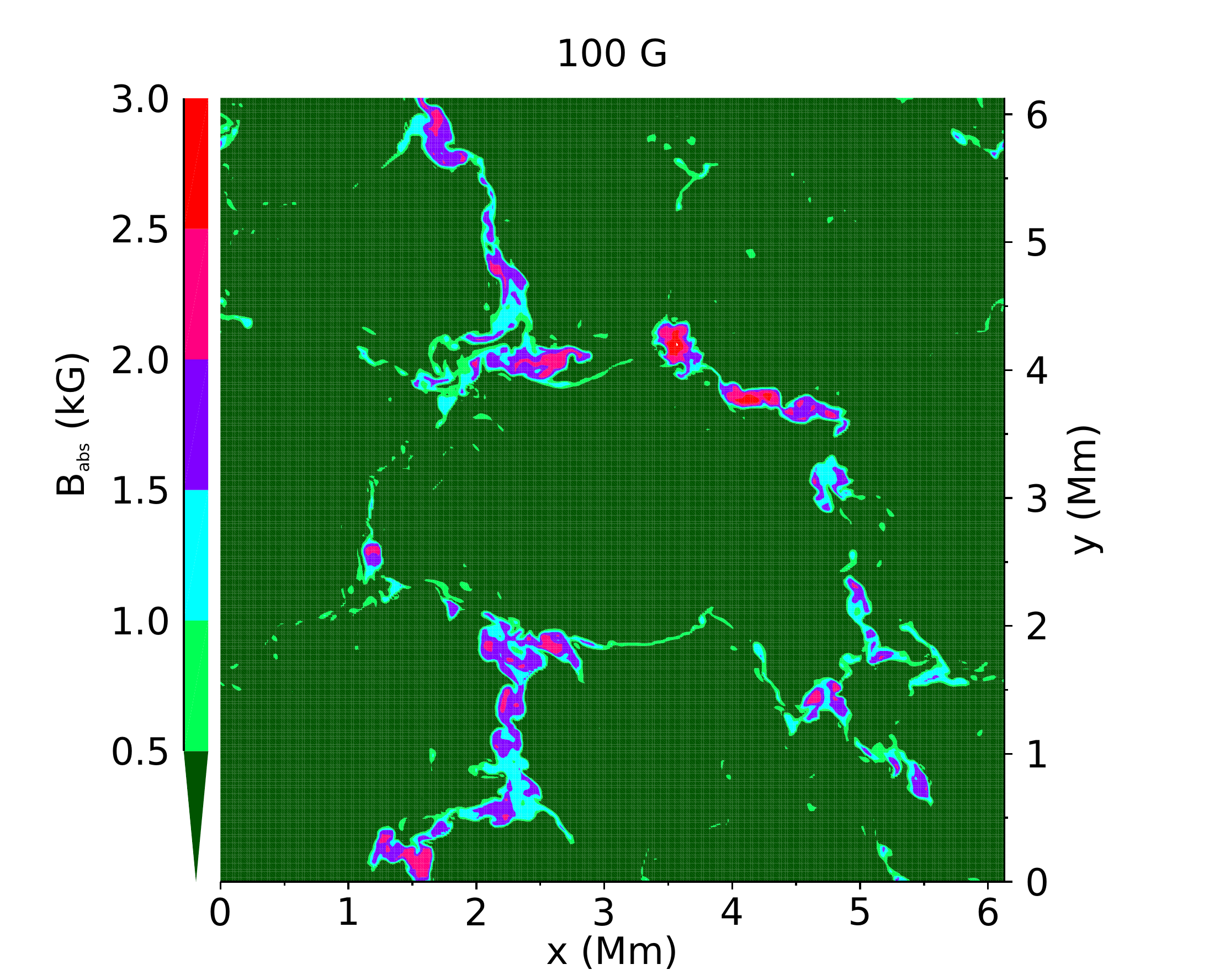}
		\hspace{0.3cm}\includegraphics[width=0.33\linewidth]{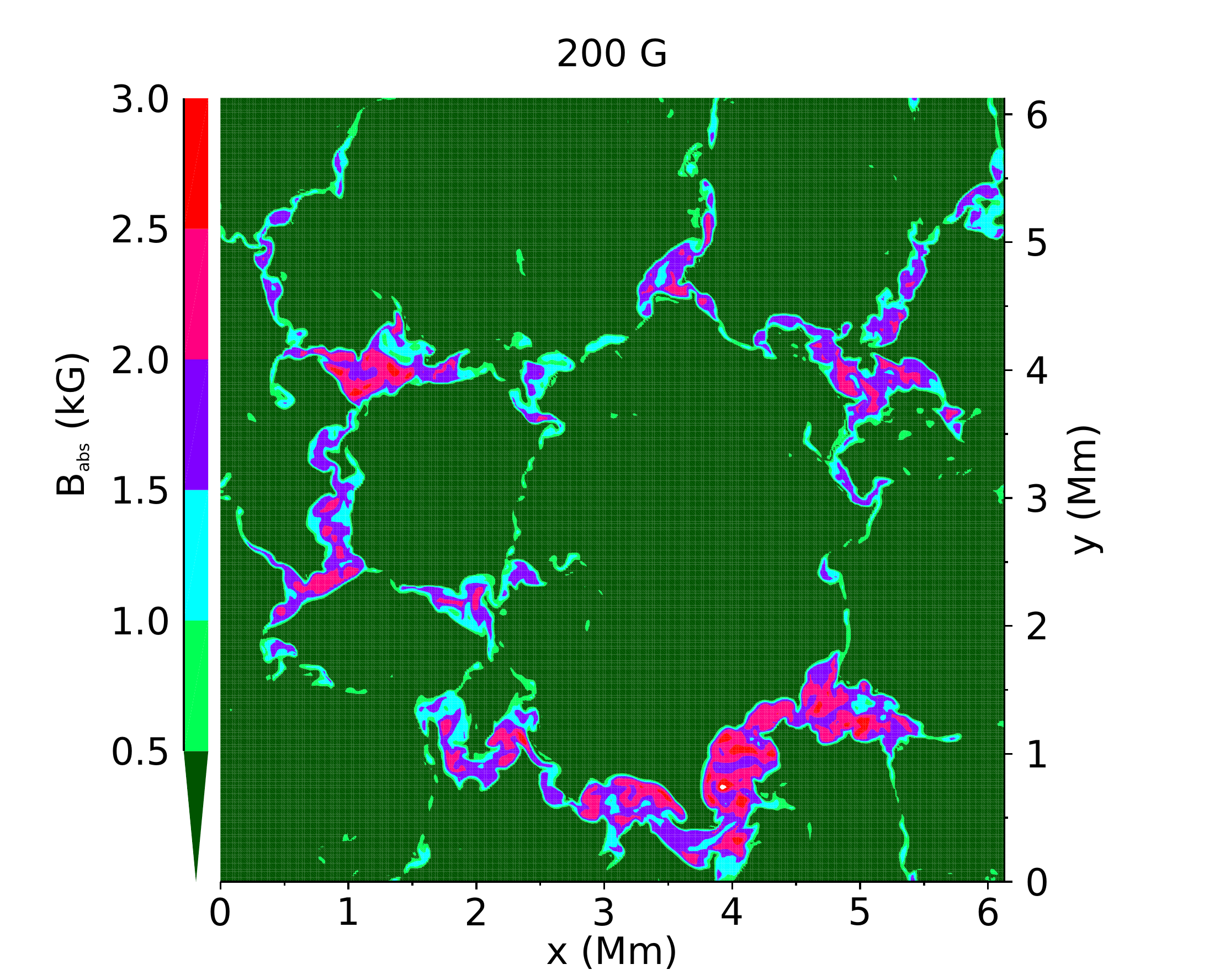}
\caption{\label{fig:Btau} Top panel: Magnetic field strengths at the height where $\tau=1$ for the 100-G and 200-G snapshots. Bottom panel: Mask of the 100-G and 200-G snapshot to identify which pixels fall into 100-G intervals of the absolute magnetic field strength over the range for each snapshot.}
	\end{center}
\end{figure*}

Second, we {make use of} the {RH} code \citep{Uitenbroek2001}. This radiative transfer code {can compute} emergent intensities at different viewing angles in different geometries. It allows the computation of several atomic and molecular transitions in both LTE and non-LTE under complete or partial redistribution. Because of its versatility,  RH is widely employed for spectral and spectro-polarimetric syntheses of atomic and molecular lines in solar and stellar atmospheres, and more recently it has been employed for solar irradiance reconstructions \citep{criscuoli2018, criscuoli2019, berrilli2020}. A {massively parallel version, RH1.5D}, allows us to compute emergent radiation on a column-by-column basis \citep{pereira2015} and has been used for syntheses {from 3D MHD simulations \citep[e.g.][]{Pereira2013, Antolin2018, Peck2019}}. {We make use of RH1.5D to allow for the efficient calculation of intensities from the simulations. Because we focus on LTE calculations for the vertically emergent intensity, there is no difference between a 1.5D calculation and a full 3D calculation.}

Third, the MURaM code is a radiative MHD code that was {originally} developed by \citet{Voegler2005}. {We use here the version of \citet{Rempel2014} that uses a different formulation of numerical diffusivities and has been tweaked for computational performance. The radiative transfer scheme is identical to \citet{Voegler2005}, but has been expanded to allow for additional diagnostic radiative transfer as described in \citet{Rempel2020}}. In this study we
use MURaM in two capacities: (1) to compute the MHD snapshots analysed and (2) to use the radiative transfer solver of MURaM in order to compute diagnostic intensities for comparison. To this end we use the approach detailed in \citet{Rempel2020} in combination with a RH based opacity table from \citet{Criscuoli2020}. 
{The MURaM RT scheme \citep{Voegler2005} uses short characteristics integration as described in \citet{KUNASZ198867}. Since gradients in opacity, density and source function are steep, MURaM uses a linear interpolation for enhanced stability. This can lead to artificial broadening of inclined rays as discussed by \citet{Peck2017SC}. Since we focus in this paper only on the intensity for vertical rays (i.e. coordinate axis aligned rays), intensity diffusion is not a concern. Finally, as already indicated above, MURaM uses the same opacities as RH.}
The main purpose of this paper is to benchmark the radiative transfer codes and to validate their performance. {For the present study to be consistent all radiative transfer calculations are done in LTE. The basic concept of the radiative transfer equations and its solution is given in the Appendix.} We are specifically interested in {validating} the codes for the continuum wavelength. We have identified the wavelength of 665.01 nm to be free of spectral lines, and {therefore adopt it for this study, hereby referring to it as 665 nm}.

\section{Results}\label{sec:results}
For each column in the MHD simulation box we have calculated the intensity spectrum at 665 nm in LTE. For the comparison of the results it turned out to be important to use an identical grid for all three radiative transfer codes. This is particularly important as the MURaM radiation scheme is inherently set up to use a grid that is shifted {from the MHD grid} by half a grid point in all three dimensions (i.e. MHD quantities are cell-centred while intensities are computed on cell corners). Data are interpolated onto this grid by using the 8-pt average of the surrounding grid cells. Figure\,\ref{fig:fringes} shows the normalized intensities for the 100-G snapshot for MURaM (left panel), RH based on the original MHD grid (middle panel), and RH using the 8-pt averaged grid (right panel). The RH intensities based on the original MHD grid show spurious fringes that disappear when using the 8-pt averaged MHD grid. We note that the decrease of the strong gradients when using interpolated atmospheres mostly results from a better sampling of the regions around the optical depth unity, where most of the radiation originates from. To confirm this point, we repeated the RH syntheses using snapshots interpolated on a vertical grid twice the resolution of the original one while keeping the original horizontal grid, and found that the intensity distribution is very similar to the one obtained on the 8-pt averaged snapshots. Overall, using interpolated grids reduces the peak intensity and increases the width of the intensity distribution, which is qualitatively what one would expect when increasing the spatial resolution.

Figure\,\ref{fig:contrast_separate1} shows normalized intensity as calculated with the MURaM radiative transfer scheme (top panels) and the RH (bottom panels). From visual inspection it is not possible to identify any difference. In Fig.\,\ref{fig:distribution}, top panels, we show the corresponding intensity distributions for all three snapshots. The {purple} lines show the intensity distribution of the RH spectral synthesis using the original MHD grid values, the red dashed lines the calculations with the 8-point interpolation of the grid cells, and the blue dotted line the MURaM intensities. 
While there are some small deviations, overall, the distributions show consistent results. Since the MHD cubes where computed with the MURaM RT scheme in the first place, and as the MHD grid interpolation removes spurious intensity fringes - while still producing consistent intensity distributions - we apply the 8-pt grid interpolation scheme for the RH and COSI calculations presented here. 

To validate the code performance in more detail, in Fig,\,\ref{fig:distribution}, bottom panels, we compare the distribution of the intensity calculated with the 8-pt interpolated snapshots for COSI (black line), RH (red dashed) and MURaM (blue dotted) obtained for the HD (left panel)), 100G (middle panel), and 200G (right panel). The comparison shows that the distributions cover the same range in absolute intensity. Furthermore, the overall shape of the distributions agrees very well, while some systematic differences can be identified. A key result of this comparison is that in the low-intensity wing of the distribution RH and MURaM agree very well, while COSI and RH reproduce the same intensities at the high-intensity wing of the distribution. {This finding is systematically present in all three snapshots. As such, it can be concluded that RH generally produces slightly wider intensity distribution than MURaM and COSI. Comparing the COSI and MURaM calculations more closely, these two distributions appear systematically shifted, with COSI producing slightly brighter intensities than MURaM. Furthermore, taking into account the shift between COSI and MURaM, the features with more abundant intensity values at the peak of the distributions are consistently reproduced in all three codes. In summary, COSI, MURaM and RH reproduce consistently the same distribution envelope as well as details with small intensity variations for all snapshots.}

We are further interested in studying where the largest differences of the intensities come from. Fig.~\ref{fig:diff} and Fig.~\ref{fig:ratio} show the difference and the ratio, respectively, between emergent intensities (normalized to their average) obtained with MURaM, RH, {and COSI}. The images indicate that in the HD snapshots differences of roughly the same amplitude between the {three} codes are found in both dark intergranular lanes and bright upflow regions. Here, the pixel-to-pixel difference between the RH versus MURaM calculations ranges from -0.04 to 0.12, and the respective ratio from 0.96 to 1.10. For COSI and the MURaM the difference ranges from -0.24 to 0.46, and the ratio between 0.77 and 1.58. 

For the 100-G snapshot, the differences between RH and MURaM range from -0.08 to 0.09 and the ratio between 0.92 to 1.10, while for the case of COSI and MURaM the difference ranges from -0.58 to 0.63, and the ratio between 0.60 and 1.73. The difference and ratio for the 200-G simulation range from -0.05 to 0.16 and 0.95 to 1.13, respectively. For COSI and MURaM the difference ranges for the 200-G simulation ranges from -0.88 to 0.56, and the ratio between 0.55 and 1.71. 
{The larger pixel-to-pixel variations of the COSI calculations compared to RH and MURaM stem from higher intensity values, mostly in the intergranular lanes, which will be further investigated.} Comparing the averages of the intensities, the ratios of RH and MURaM give 1.012, 1.013, and 1.014  {and for COSI and MURaM it is 1.042, 1.041, and 1.040 for the HD, 100-G, and 200-G snapshots, respectively}. From this we conclude that while the pixel-to-pixel differences of the codes can vary by several percent. The averaged snapshots agree to 1.4\,\% for {the codes using the same opacities and to about 3-4\,\% for the codes with different opacity sets}.

{Understanding how the intensity scales with the magnetic field strength is especially important in the context of irradiance studies, as some irradiance reconstruction models make use of the magnetic flux as proxy for the radiative intensity \citep[e.g.][]{krivova2003, foukal2011,yeo2017b}}. As such, the relation between brightness and magnetic flux has been the subject of several observational \citep[e.g.][]{ortiz2002,criscuoli2017} and theoretical studies \citep[e.g.][]{rohrbein2011, Peck2019}. 
\begin{figure}[t!]
	\begin{center}
    \includegraphics[width=0.95\linewidth]{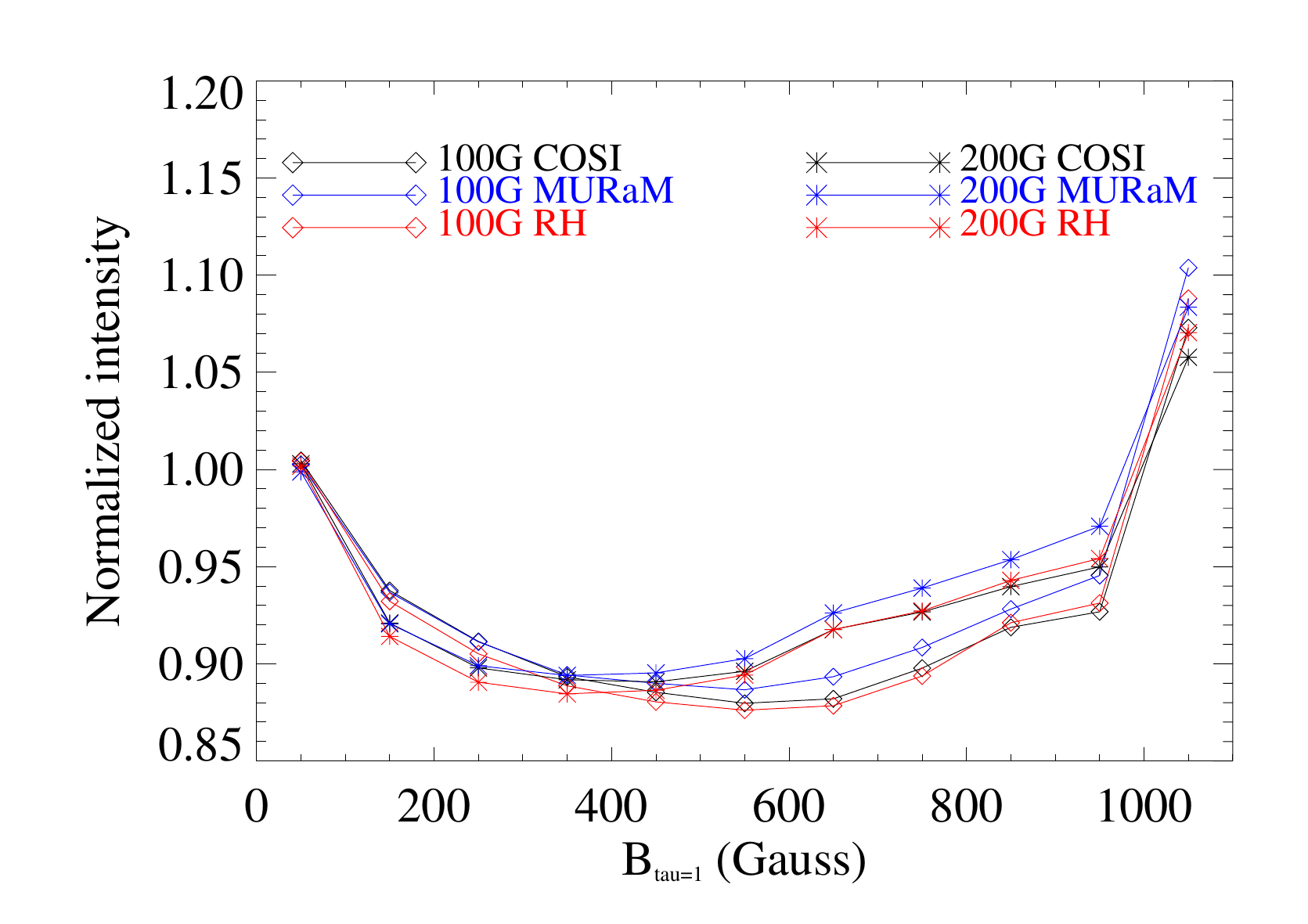} \\
\caption{\label{fig:Int_B} Normalized intensity as a function of absolute magnetic field strength $B_{\tau=1}$ for COSI (black), MURaM  (blue), and RH (red). The diamonds give the normalized intensity determined from the 100-G simulation, while the stars the ones from the 200-G run. The last bin contains all magnetic field strength > 1000\,G.}
\end{center}
\end{figure}
Therefore, we are further interested {in finding} how well the intensities calculated by the different codes agree for different magnetic field strengths. Figure\,\ref{fig:Btau}, top panels, shows the magnetic field strengths for each snapshot, interpolated to the height where $\tau=1$. The bottom panels show the respective absolute field strength for both snapshots for the same layer { segmented at 500G intervals }. We then investigate how the normalized intensity depends on absolute magnetic field strength at $\tau=1$ for subsequent 100-G bins in both snapshots. Figure\,\ref{fig:Int_B} compares the mean intensity for each of the bins for the 100-G and 200-G snapshots calculated with COSI (black), MURaM (blue) and RH (red). We note that the shape of these curves is heavily influenced by realization noise, since they are based on single MHD snapshots { (see \citet{Rempel2020} for a detailed discussion on the influence of realization noise on the emergent intensity. Consequently,} while 
we cannot {meaningfully} compare the differences between the 100 and 200 G snapshots, we can compare the differences of the 3 radiative transfer codes for each of the setups, { which is the goal of this work}. 
Nonetheless, the overall shape of the dependence of the intensity as a function of magnetic field strength is in line with the findings by \cite{Rempel2020}. Specifically, the intensities in regions with small-medium magnetic field strength (100-900 G) are darker than in the non-magnetic areas. Second, the shape of the intensity variation as a function of magnetic field strength shows systematic differences for the 100-G and 200-G snapshots. In particular, the turning point of the intensity in the 100-G snapshot is found at around 500\,G, while for the 200-G snapshot it is at around 350\,G. This difference can be explained by the fact that the stronger magnetic field in the 200-G snapshot pushes the weak field, which is naturally confined to the intergranular lanes, towards {the edges of granular cells} and as such to areas with higher intensity. Third, while all three codes agree very well in their response to the varying plasma properties some detailed performance differences can still be identified. The scatter amongst the codes is systematically lower for the 100-G than for the 200-G snapshots. This goes in line with the differences found in Fig.\,\ref{fig:diff}, where the RH and MURaM code show the largest deviations in difference (top panels) and ratio (bottom panels) for the 200-G snapshot, and specifically in the intergranular lanes. Looking at the codes individually, in the 100-G snapshots all three codes agree within almost about 1-2\,\% and the codes do not seem to produce systematic differences for that individual snapshot. This is somewhat different for the 200-G snapshot. Here the codes differ by about 2-3\,\%. Moreover, the MURaM code tends to give systematically brighter intensity for the weak-medium magnetic field strengths. An indication of that can be also found in Fig.\,\ref{fig:distribution}, top panel, where the MURaM calculations give consistently higher intensities at the low-intensity side of the peak of the distributions than the RH code. Finally, the average intensities for all magnetic field strength larger than 1000 G lead to a normalized intensity above unity.
\section{Conclusions}\label{sec:concl}
We have determined the emergent intensity from 3D MHD simulation snapshots for a non-magnetic case, and 100 G and 200 G simulation using the radiative transfer codes COSI, RH and the MURaM radiation scheme. We compared the difference, ratio and distribution of the intensities and find an overall good agreement amongst the codes. {We find that while the absolute intensities produced by the codes agree very well, systematic differences on the code performance could be identified from the distribution functions. In particular, the RH code produces slightly wider distributions. The RH and MURaM calculations agree very well in the low-intensity range, while RH and COSI match very well on the higher intensities.} While the pixel-to-pixel differences of the codes can vary by several percent, the averaged intensities for RH versus MURaM differ up to 1.4\,\% and COSI versus RH and MURaM to about 3-4\,\%, respectively. 

Furthermore, we investigated how well the codes reproduce the intensities for different magnetic field strengths at the $\tau$=1 layer. Here, we find the differences between the codes to be about 1-2\,\% for the 100-G snapshot and about 2-3\,\% for the 200-G snapshot. The overall shape of the change in intensity as a function of magnetic field strength is in line with previous work. For the COSI and RH calculations we carried out an 8-point averaging of the cell corners of the MHD grid onto the cell centers as it is done in the MURaM code. In addition to ensuring consistency, this technique also removes fringes that appear when using the original MHD grid with RH and COSI. While we focused in this study on computing intensities, it is likely that this technique should also be considered for polarized radiative transfer as well. Tests performed using snapshots interpolated on the vertical grid suggest that MHD simulations with substantially higher resolution in the vertical $\tau$-scale might not cause this issue. A detailed study investigating resolution dependence would need to follow.
\begin{acknowledgements}
 We kindly acknowledge the support by the International Space Science Institute (ISSI), Bern, Switzerland during the International Team on Modeling Solar Irradiance with 3D MHD simulations, lead by Serena Criscuoli. MH kindly acknowledges support by Daniel Karbacher. TMDP's research is supported by the Research Council of Norway through its Centres of Excellence scheme, project number 262622. The National Solar Observatory is operated by the Association of Universities for Research in Astronomy, Inc. (AURA)
under cooperative agreement with the National Science Foundation. This material is based upon work supported by the National Center for Atmospheric Research, which is a major facility sponsored by the National Science Foundation under Cooperative Agreement No. 1852977.
\end{acknowledgements}

\begin{appendix}
\section{Basic concept of radiative transfer}
In the following we describe the basic concept of how the radative transfer is solved in the case of local thermodynamic equilibrium and then discuss the differences in the implementation in the case of MURaM, RH and COSI.

In local thermodynamic equilibrium (LTE) the velocity distribution of the atoms, ions and electrons is Maxwellian. Furthermore, the ionization rates and population numbers are a function of the local temperature and density. The assumption of LTE is only true if the collisional processes dominate over the radiative processes, e.g. for high densities, which is a suitable assumption for our study. The standard radiative transport equation in Eq.\,\ref{equ:transp}
\begin{equation}
\centering
\mu \frac{dI_{\nu\mu}}{dz} = \eta_{\nu} -\chi_{\nu} I_{\nu\mu}
\label{equ:transp}
\end{equation}
describes the change of the specific intensity $I_{\nu}$ at frequency $\nu$ along the path $dz$ in a plane-parallel atmosphere due to the emission and absorption, where $\eta$ is the emissivity and $\chi$ the extinction coefficient. In our study $\mu = \cos\Theta$=1 as we consider the radiative transfer in the vertical direction. Using the optical depth
\begin{equation}
\centering
\tau_{\nu} = \int_0^{z'}\chi_{\nu} dz, \label{equ:tau}
\end{equation}
and the source function, defined as
\begin{equation}
\centering
S_{\nu}  \equiv \frac{\eta_{\nu}}{\chi_{\nu}}, \label{equ:source}
\end{equation}
the intensity change can be described as a function of the change of the optical depth as
\begin{equation}
\centering
\frac{dI_{\nu}}{d\tau}  = I(d\tau_\nu) - S(d\tau_\nu). \label{equ:inttau}
\end{equation}

\noindent Multiplying Eq.\,\ref{equ:inttau} with $e^{-\tau}$ gives
\begin{equation}
\centering
\frac{d I_{\nu} e^{-\tau_\nu}}{d\tau_{\nu}} = I_{\nu} e^{-\tau} - S_{\nu} e^{-\tau_\nu} \label{eq:fac}
\end{equation}

\noindent If $S_{\nu}$ is known, the integration of Eq.\,\ref{eq:fac} gives
\begin{equation}
\centering
I_{\nu}(\tau_1) = I_{\nu}(\tau_2) e^{-(\tau_2-\tau_1)} + \int_{\tau_1}^{\tau_2}S_{\nu}(t)e^{-(t)} dt, \label{equ:solution}
\end{equation}

\noindent It is assumed $S_\nu$ is a linear function of $\tau_\nu$. Then Eq.\,\ref{equ:solution} can be written as:
\begin{equation}\label{eq:formal}
\centering
I_{\nu}(\tau_1) = I_{\nu}(\tau_2) e^{-(\tau_2-\tau_1)} + \int_{\tau_1}^{\tau_2} e^{-t}[S_1+S_2(t-\tau_1)/(\tau_2-\tau_1)] dt
\end{equation}
Clearly, the determination of the opacities and the source function has an impact in the emergent intensity. In the following, we give the basic concept of the numerical scheme of all three codes that is used to determine the emergent intensity in LTE and vertical direction ($\mu=1$).
\section{MURaM}
MURaM interpolates the opacity from a RH opacity table $\kappa=f(\rho, T)$ using a bi-linear table interpolation. Starting with the values at the positions: $z_1$ ($\kappa_1$, $\varrho_1$, $S_1$) and $z_2$  ($\kappa_2$, $\varrho_2$, $S_2$) we compute the $\tau$ scale using $\Delta \tau=\tau_2-\tau_1>0$:
\begin{equation}
    \Delta \tau=-\Delta z [(\kappa_1 \varrho_1+\kappa_2 \varrho_2)/3+(\kappa_1 \varrho_2+\kappa_2 \varrho_1)/6]
\end{equation}
The outgoing intensity at the top of the domain is computed solely for diagnostics using a scheme that is separate from the radiative transfer that is used to compute the intensity throughout the simulation domain (needed for radiative heating/cooling). The contribution to the outgoing intensity at the top of the domain from the interval $[\tau_1, \tau_2]$ in regions with $\Delta\tau>10^{-5}$ is given by:
\begin{eqnarray}
    c&=& (1-e^{-\Delta \tau})/\Delta \tau\\
    \Delta I &=& [S_1 (1-c) + S_2 (c-e^{-\Delta \tau})] e^{-\tau_1}
\end{eqnarray}
otherwise:
\begin{eqnarray}
    \Delta I &=& 0.5\Delta\tau[S_1 + S_2] e^{-\tau_1}
\end{eqnarray}
\section{RH}
In RH, the formal solution of Eq. (\ref{equ:solution}), is achieved by piecewise integration using the formalism of characteristics \citep[see][]{KUNASZ198867}. In this work we used the solver with linear interpolation of the source function between points $\tau_1$ and $\tau_2$, according to the expression:

\begin{equation}
    I_\nu(\tau_1) = I_\nu(\tau_2) e^{-\Delta\tau} + w_0 S_1 + w_1 \frac{\Delta S}{\Delta \tau},
\end{equation}
where
\begin{eqnarray*}
\Delta\tau & = & 0.5 \Delta z (\chi_1 + \chi_2),\\
\Delta S &=& S_2 - S_1,\\
w_0 & = & 1 - e^{-\Delta\tau},\\
w_1 & = & w_0 - \Delta\tau e^{-\Delta\tau}.
\end{eqnarray*}
To preserve numerical accuracy, when $\Delta\tau > 50$, $w_0=w_1=1$ and when $\Delta\tau < 5\cdot10^{-4}$, $w_0$ and $w_1$ are approximated by
\begin{eqnarray*}
w_0 & = & \Delta\tau \left(1 - \frac{\Delta\tau}{2}\right),\\
w_1 & = &  \Delta\tau^2 \left(\frac{1}{2} - \frac{\Delta\tau}{3}\right).
\end{eqnarray*}

In RH the extinction coefficients are explicitly calculated, while in MURaM they are interpolated from the RH opacity table ($\kappa\equiv\chi/\rho=f(\rho, T)$).
For consistency, both the explicit RH synthesis and the opacity table employed in MURaM were computed under LTE using the same set of input parameters. In summary, the synthesis included the full list of atomic and molecular bound-free transitions from the Kurucz website, photo-ionization of 12 atomic species (including updated Fe atom cross-sections) and photo-dissociation of 52 diatomic molecules. The computation took also into account  Thomson and Rayleigh scattering, although scattering contribution is expected to be negligible in continua along the vertical line-of-sight \citep[e.g.][]{Fabbian2015}.  
\section{COSI}
The numerical scheme of COSI consists of two modules. The first, also referred to as {\it hminus} \citep{Haberreiter2008b,shapiro2010} calculates the level populations of the atoms, which under the assumption of LTE follow the Boltzmann distribution
\begin{equation}
\frac{n_{\mathrm{u}}}{n_{\mathrm{l}}} = \frac{g_{\mathrm{u}}}{g_{\mathrm{l}}}e^{-(\chi_{\mathrm{u}}-\chi_{\mathrm{l}})/kT} = \frac{g_{u}}{g_{l}}e^{-(h\nu/kT)}.   \label{equ:boltz}
\end{equation}

The second module of the code, also referred to as {\it fioss}, uses the these level populations to calculate the emergent intensity, in our case in the vertical direction ($\mu=1$). In COSI the opacities are determined taking into account all radiative (negligible in the case of LTE) and collisional processes for absorption, emission and ionization. In COSI the solution of Eq.\,(\ref{equ:solution}) is obtained via consideration of the change of the intensity between 3 subsequent depth points, $n$=1,2,3 and $\tau_3 > \tau_2 > \tau_1$:\\
if $\tau > 10^{-10}$ then
\begin{eqnarray}
I_{\nu} &=&I_{\nu}e^{-\Delta \tau} + dS_{\nu}/d\tau =\\
 &=&I_{\nu}e^{-\Delta \tau} + S_{\nu}w_\tau
\end{eqnarray}
with
\begin{eqnarray}
w_{\tau} & = &  w_{a}+w_{b}\\
w_{a} & = & (e^{-\tau_1} - e^{-\tau_2})/\Delta \tau_1 \\
w_{b} & = & (e^{-\tau_3} - e^{-\tau_2})/\Delta \tau_2 \\
\Delta\tau_1&=&0.5 \,\Delta z_1 (\chi_1+\chi_2) \\
\Delta\tau_2&=&0.5\,\Delta z_2(\chi_2+\chi_3). 
\end{eqnarray}
Note that the $S_2 e^{-\tau_2}$ term (from equation ) cancels for consecutive depth points with the next term, which is then equal to the new $S_1$-term.
\end{appendix}

%
%

\end{document}